\documentclass[11pt]{article}

\usepackage{amscd}
\usepackage{amsmath}
\begin{document}

\newcommand{\N}{N\raise.7ex\hbox{\underline{$\circ $}}$\;$}

\thispagestyle{plain}

\title{Majorana-Oppenheimer approach to Maxwell \\ electrodynamics
in Riemannian space-time }
\author{  Bogush A., Red'kov V.,    Tokarevskaya N.,  Spix G.  }

\date{}

\maketitle

\begin{abstract}

The Riemann -- Silberstein -- Majorana -- Oppengeimer approach to
the Maxwell electro\-dyna\-mics in presence of electrical sources
and arbitrary media is investigated within the  matrix formalism.
The symmetry of the matrix Maxwell equation under transformations
of the complex rotation group SO(3.C) is demonstrated explicitly.
In vacuum case, the matrix form includes  four  real $4 \times 4$
matrices $\alpha^{b}$. In presence of media matrix form requires
two  sets of   $4 \times 4$ matrices, $\alpha^{b}$ and $\beta^{b}$
-- simple and symmetrical realization of which is given. Relation
of  $\alpha^{b}$ and $\beta^{b}$  to   the Dirac matrices in
spinor basis is found. Minkowski constitutive relations in case of
any linear media are given in a short algebraic
 form based on the use  of complex 3-vector fields and
complex orthogonal rotations from SO(3.C) group. The  matrix
complex formulation in  the Esposito's form, based on the use of
two electromagnetic 4-vector, is studied and discussed. Extension
of the 3-vector complex  matrix formalism to arbitrary  Riemannian
space-time in accordance with tetrad method by
Tetrode-Weyl-Fock-Ivanenko is performed.
\end{abstract}

\section{Introduction}

Special relativity arose from study of the symmetry properties of
the Maxwell equations with respect to motion of references frames:
Lorentz  \cite{1904-Lorentz},  Poincar'e \cite{1905-Poincare},
Einstein \cite{1905-Einstein} Naturally, an analysis of the
Maxwell equations with respect to Lorentz transformations was the
first objects of relativity theory:
 Minkowski \cite{1908-Minkowski},
 Silberstein \cite{1907-Silberstein(1)},\cite{1907-Silberstein(2)},
Marcolongo \cite{1914-Marcolongo},  Bateman  \cite{1915-Bateman},
and  Lanczos \cite{1919-Lanczos}, Gordon \cite{1923-Gordon},
Mandel'stam --  Tamm
\cite{1925-Mandel'stam},\cite{1925-Tamm(1)},\cite{1925-Tamm(2)}.

After discovering the relativistic equation for a particle with
spin 1/2 -- Dirac \cite{1928-Dirac} -- much work was done to study
spinor and vectors within the Lorentz group theory: M\"{o}glich
\cite{1928-Moglich},  Ivanenko -- Landau
\cite{1928-Ivanenko-Landau}, Neumann \cite{1929-Neumann}, van der
Waerden \cite{1929-Waerden}, Juvet \cite{1930-Juvet}. As was shown
any quantity which transforms linearly under Lorentz
transformations is a spinor. For that reason spinor quantities are
considered as fundamental in quantum field theory and basic
equations for such quantities should be written in a spinor form.
A spinor formulation of Maxwell equations was studied by Laporte
and Uhlenbeck \cite{1931-Laporte}, also see Rumer
\cite{1936-Rumer}. In 1931,   Majorana \cite{1931-Majorana} and
Oppenheimer \cite{1931-Oppenheimer} proposed to consider the
Maxwell theory of electromagnetism as the wave mechanics of the
photon. They introduced a  complex 3-vector wave function
satisfying the massless Dirac-like equations. Before Majorana  and
Oppenheimer,    the most crucial steps were made by Silberstein
\cite{1907-Silberstein(1)}, he showed the possibility to have
formulated Maxwell equation in term of complex 3-vector entities.
Silberstein in his second paper  \cite{1907-Silberstein(2)} writes
that the complex form of Maxwell equations has been known before;
he refers there to the second volume of the lecture notes on the
differential equations of mathematical physics by B. Riemann that
were edited and published by H. Weber in 1901 \cite{1901-Weber}.
 This not widely used fact  is  noted by Bialynicki-Birula   \cite{1994-Bialynicki-Birula}).

Maxwell equations in the  matrix Dirac-like  form considered
during long time by many authors, the interest to the
Majorana-Oppenheimer formulation of electrodynamics has grown in
recent years:

Luis de Broglie
\cite{1934-Broglie(1)},\cite{1934-Broglie(2)},\cite{1939-Broglie},\cite{1940-Broglie},
Mercier \cite{1935-Mercier},
 Petiau \cite{1936-Petiau}, Proca \cite{1936-Proca}, \cite{1946-Proca},
Duffin \cite{1938-Duffin}, Kemmer
\cite{1939-Kemmer},\cite{1943-Kemmer},\cite{1960-Kemmer},  Bhabha
\cite{1939-Bhabha}, Belinfante
\cite{1939-Belinfante(1)},\cite{1939-Belinfante(2)}, Taub
\cite{1939-Taub},  Sakata  -- Taketani \cite{1940-Sakata},
Schr\"{o}dinger  \cite{1940-Schrodinger},
\cite{1943-Schrodinger(1)},\cite{1943-Schrodinger(2)}, Tonnelat
cite{1941-Tonnelat}, Heitler \cite{1943-Heitler},
\cite{1946-Harish-Chandra(1)},\cite{1946-Harish-Chandra(2)},
Hoffmann \cite{1947-Hoffmann}, Utiyama \cite{1947-Utiyama},
Mercier \cite{1949-Mercier}, Imaeda \cite{1950-Imaeda},   Fujiwara
\cite{1953-Fujiwara}, G\"{u}rsey \cite{1954-Gursey}, Gupta
\cite{1954-Gupta}, Lichnerowicz \cite{1954-Lichnerowicz}, , Ohmura
\cite{1956-Ohmura},
  Borgardt \cite{1956-Borgardt},\cite{1958-Borgardt},  Fedorov \cite{1957-Fedorov},  Kuohsien \cite{1957-Kuohsien},
Bludman   \cite{1957-Bludman}, Good \cite{1957-Good},  Moses
\cite{1958-Moses}-\cite{1959-Moses}-\cite{1973-Moses},  Silveira
\cite{1959-Silveira}, \cite{1980-Silveira}, Lomont
\cite{1958-Lomont}, Kibble \cite{1961-Kibble}, Post
\cite{1961-Post}, Bogush -- Fedorov \cite{1962-Bogush-Fedorov},
Sachs -- Schwebel  \cite{1962-Sachs-Schwebel},   Ellis
\cite{1964-Ellis}, Oliver \cite{1968-Oliver}, Beckers -- Pirotte
\cite{1968-Beckers},  Casanova \cite{1969-Casanova},  Carmeli
\cite{1969-Carmeli}, Bogush \cite{1971-Bogush}, Lord
\cite{1972-Lord}, Weingarten \cite{1973-Weingarten},   Mignani --
Recami --  Baldo  \cite{1974-Recami1}, \cite{1974-Frankel},
\cite{1975-Jackson}, Edmonds \cite{1975-Edmonds},  Strazhev --
Tomil'chik  \cite{1975-Strazhev-Tomil'chik}, Silveira
\cite{1980-Silveira},   Jena -- Naik --  Pradhan \cite{1980-Jena},
Venuri  \cite{1981-Venuri}, Chow \cite{1981-Chow}, Fushchich --
Nikitin \cite{1983-Fushchich}, Cook
\cite{1982-Cook(1)}-\cite{1982-Cook(2)}, Giannetto
\cite{1985-Giannetto},   -- Y\'epez, Brito -- Vargas
\cite{1988-Nunez}, Kidd --  Ardini --  Anton  \cite{1989-Kidd},
Recami \cite{1990-Recami}, Krivsky --  Simulik
\cite{1992-Krivsky},  Hillion \cite{1993-Hillion}, Baylis
\cite{1993-Baylis},
 Inagaki \cite{1994-Inagaki},
Bialynicki-Birula
\cite{1994-Bialynicki-Birula}-\cite{1996-Bialynicki-Birula}-\cite{2005-Birula},
Sipe \cite{1995-Sipe}, \cite{1996-Ghose}, Esposito
\cite{1998-Esposito}, Dvoeglazov  \cite{1998-Dvoeglazov'},
\cite{1998-Dvoeglazov} (see a big list of  relevant references
therein)-\cite{2001-Dvoeglazov}, Gersten  \cite{1998-Gersten},
Gsponer \cite{2002-Gsponer},
\cite{2001-Ivezic(1)},\cite{2002-Ivezic},\cite{2002-Ivezic(2)},\cite{2002-Ivezic(3)},\cite{2003-Ivezic},
\cite{2005-Ivezic(1)},\cite{2005-Ivezic(2)},\cite{2005-Ivezic(3)},\cite{2006-Ivezic},
Donev --  Tash\-kova \cite{2004-Donev(1)}, \cite{2004-Donev(2)},
\cite{2004-Donev(3)}, Armour \cite{2004-Armour}.

Our treatment will be with a quite definite  accent:
 the main attention is given to technical aspect of classical electrodynamics
 based on the theory of rotation complex group SO(3.C) (isomorphic to the Lorentz group -- see
 Kur\c{s}uno$\hat{g}$lu \cite{1961-Kursunoglu},
Macfarlane  cite\cite{1962-Macfarlane}-\cite{1966-Macfarlane},
Fedorov \cite{1979-Fedorov}).

\section{  Complex matrix form of Maxwell theory in vacuum }

Let us start with Maxwell equations in  a uniform ($\epsilon,
\mu$)-media  in  presence  of  external sources
\cite{1941-Stratton}-\cite{1961-Panofsky-Phillips}-\cite{1975-Jackson}:
\begin{eqnarray}
 \mbox{div} \; c{\bf B} = 0 \; , \qquad \mbox{rot}
\;{\bf E} = -{\partial c {\bf B} \over \partial ct}  \nonumber
\\
 \mbox{div}\; {\bf E} = {\rho \over \epsilon
\epsilon_{0}} , \qquad
 \mbox{rot} \; c{\bf B} = \mu \mu_{0}c {\bf J} + \epsilon \mu
   {\partial {\bf E} \over \partial ct} \; .
\label{1.1a}
\end{eqnarray}

\noindent With the use of usual notation for  current 4-vector $
j^{a} = (\rho, {\bf J} /c) \; ,  \; c^{2} = 1 /
\epsilon_{0}\mu_{0} \; , $  eqs. (\ref{1.1a}) read (first,
consider the  vacuum case):
\begin{eqnarray}
 \mbox{div} \; c{\bf B} = 0 \; , \qquad \mbox{rot}
\;{\bf E} = -{\partial c {\bf B} \over \partial ct}  \nonumber
\\
 \mbox{div}\; {\bf E} = {\rho \over  \epsilon_{0}} , \qquad
 \mbox{rot} \; c{\bf B} =  {{\bf j} \over \epsilon_{0}}  +
   {\partial {\bf E} \over \partial ct} \; ,
\label{1.2a}
\end{eqnarray}

Let us introduce 3-dimensional complex vector $\psi^{k} =   E^{k}
+ i c B^{k} $,
 with the help of which the above equations can be  combined into (see  Zilberschtein  \cite{1907-Silberstein(1)}-\cite{1907-Silberstein(2)},
 Bateman  \cite{1915-Bateman}, Majorana \cite{1931-Majorana},  Oppenheimer \cite{1931-Oppenheimer},
 and many others)
\begin{eqnarray}
\partial_{1}\Psi ^{1} + \partial_{2}\Psi ^{0} + \partial_{3}\Psi ^{3} =
j^{0} / \epsilon_{0}  \; ,\;\; -i\partial_{0} \psi^{1} +
(\partial_{2}\psi^{3} -
\partial_{3}\psi^{2}) = i\; j^{1} / \epsilon_{0}
\nonumber
\\
-i\partial_{0} \psi^{2} + (\partial_{3}\psi^{1} -
\partial_{1}\psi^{3}) = i\; j^{2} / \epsilon_{0} \; , \;
-i\partial_{0} \psi^{3} + (\partial_{1}\psi^{2} -
\partial_{2}\psi^{1}) =  i\; j^{3} / \epsilon_{0} \; .
\nonumber
\end{eqnarray}

\noindent  let $x_{0}=ct, \; \partial_{0} = c \;\partial_{t}$.
These four relations can be rewritten in a  matrix form using a
4-dimensional column  $\Psi$ with one additional zero-element
[Fuschich -- Nikitin \cite{1983-Fushchich}:
\begin{eqnarray}
(-i\alpha^{0} \partial_{0} + \alpha^{j} \partial_{j} ) \Psi = J \;
, \qquad \Psi = \left | \begin{array}{c} 0 \\\psi^{1} \\\psi^{2}
\\ \psi^{3}
\end{array} \right | \; , \qquad
\alpha^{0} = \left | \begin{array}{rrrr}
a_{0} & 0  &  0  & 0  \\
a_{1} & 1  &  0  & 0  \\
a_{2} & 0  &  1  & 0  \\
a_{3} & 0  &  0  & 1
\end{array}  \right |
\nonumber
\\
\alpha^{1} = \left | \begin{array}{rrrr}
b_{0} & 1  &  0  & 0  \\
b_{1} & 0  &  0  & 0  \\
b_{2} & 0  &  0  & -1  \\
b_{3} & 0  &  1  & 0
\end{array}  \right |  ,
\alpha^{2} = \left | \begin{array}{rrrr}
c_{0} & 0  &  1  & 0  \\
c_{1} & 0  &  0  & 1  \\
c_{2} & 0  &  0  & 0  \\
c_{3} & -1  & 0  & 0
\end{array}  \right |  ,
\alpha^{3} = \left | \begin{array}{rrrr}
d_{0} & 0  &  0  & 1  \\
d_{1} & 0  & -1  & 0  \\
d_{2} & 1  &  0  & 0  \\
d_{3} & 0  &  0  & 0
\end{array}  \right |  .
\nonumber
\end{eqnarray}

Here, there arise four ambiguously determined matrices; numerical
parameters  $a_{k}, b_{k}, c_{k}, d_{k}$ are arbitrary. Our choice
for the matrix form of eight  Maxwell equations  is the following:
\begin{eqnarray}
(-i \partial_{0} + \alpha^{j} \partial_{j} ) \Psi =J \; , \qquad
\Psi = \left | \begin{array}{c} 0 \\\psi^{1} \\\psi^{2} \\
\psi^{3}
\end{array} \right | \; , \qquad J=
{1 \over \epsilon_{0}} \; \left | \begin{array}{c} j^{0} \\ i\;
j^{1} \\ i\; j^{2} \\ i \; j^{3}
\end{array} \right |
\label{1.10} \end{eqnarray}

\noindent where
\begin{eqnarray}
 \alpha^{1} = \left |
\begin{array}{rrrr}
0 & 1  &  0  & 0  \\
-1 & 0  &  0  & 0  \\
0 & 0  &  0  & -1  \\
0 & 0  &  1  & 0
\end{array}  \right |,
\alpha^{2} = \left | \begin{array}{rrrr}
0 & 0  &  1  & 0  \\
0 & 0  &  0  & 1  \\
-1 & 0  &  0  & 0  \\
0 & -1  & 0  & 0
\end{array}  \right |,
\alpha^{3} = \left | \begin{array}{rrrr}
0 & 0  &  0  & 1  \\
0 & 0  & -1  & 0  \\
0 & 1  &  0  & 0  \\
-1 & 0  &  0  & 0
\end{array}  \right |
\nonumber
\\
(\alpha^{1})^{2} = -I , \qquad   (\alpha^{2})^{2} = -I ,  \qquad
(\alpha^{2})^{2} = -I  \nonumber
\\
\alpha^{1} \alpha^{2}= - \alpha^{2} \alpha^{1} =  \alpha^{3} \;,
\qquad  \alpha^{2} \alpha^{3} = - \alpha^{3} \alpha^{2} =
\alpha^{1}\;, \qquad  \alpha^{3} \alpha^{1} = - \alpha^{1}
\alpha^{3} = \alpha^{2}\;. \nonumber
\end{eqnarray}

Let us  consider the problem of relativistic invariance of this
equation. The lack of manifest invariance of 3-vector complex form
of Maxwell theory has been intensively discussed  in various
aspects: for instance, see  Esposito \cite{1998-Esposito}, Ivezic
\cite{2001-Ivezic(1)}, \cite{2002-Ivezic}, \cite{2002-Ivezic(2)},
\cite{2002-Ivezic(3)}, \cite{2003-Ivezic}, \cite{2005-Ivezic(1)},
\cite{2005-Ivezic(2)}, \cite{2005-Ivezic(3)}, \cite{2006-Ivezic}).
Let  us start with relations:
\begin{eqnarray}
(-i \partial_{0} + \alpha^{j} \partial_{j} ) \Psi =J \; , \qquad
\Psi = \left | \begin{array}{c} 0 \\\psi^{1} \\\psi^{2} \\
\psi^{3}
\end{array} \right | \;, \qquad
J = {1 \over \epsilon_{0}} \; \left | \begin{array}{c} j^{0} \\
i\; j^{1} \\ i\; j^{2} \\ i \; j^{3}
\end{array} \right | .
\nonumber
\end{eqnarray}

Arbitrary  Lorentz transformation over the function $\Psi$ is
given by (take notice that one may introduce four undefined
parameters $s_{0}, ...,s_{3}$, but we will take $s_{0}=1, s_{j}=0$
)
\begin{eqnarray}
S = \left | \begin{array}{cccc}
s_{0} & 0 & 0 & 0  \\
s_{1} & .  & . & . \\
s_{2} & .   &O(k) & . \\
s_{3} & . & . & .
\end{array} \right | \;, \qquad \Psi' = S \Psi \; , \qquad  \Psi = S^{-1} \Psi '\; ,
\label{1.13a}
\end{eqnarray}

\noindent where $O(k)$ stands for a $(3\times 3$-rotation complex
matrix from $SO(3,C)$, isomorphic to the Lorentz  group -- more
detail see in \cite{1979-Fedorov} and below in the present text.
Equation for a primed function  $\Psi'$  is
\begin{eqnarray}
(-i \partial_{0} + S \alpha^{j}S^{-1}  \partial_{j} ) \; \Psi '=
S\;J \; . \nonumber
\end{eqnarray}

\noindent When working with matrices $\alpha^{j}$ we will use
vectors
 ${\bf e}_{i}$ and ($ 3 \times 3$)-matrices $\tau_{i}$, then the structure $S \alpha^{j}S^{-1} $ is
\begin{eqnarray}
S \alpha^{j}S^{-1} =  \left | \begin{array}{cc}
0 & {\bf e}_{j} O^{-1}(k)   \\
- O(k){\bf e}^{t}_{j} &  O(k) \tau_{j} O^{-1}(k)
\end{array} \right |  = \alpha^{m} O_{mj}(k) \; .
\label{1.13c}
\end{eqnarray}

\noindent Therefore,   the matrix Maxwell equation becomes
\begin{eqnarray}
 (-i \partial_{0} + \alpha^{m}\;  \partial'_{m} ) \Psi'  = S J \;, \qquad
 O_{mj}   \partial_{j} = \partial_{m}' \;  .
\label{1.13d}
\end{eqnarray}

Now, one  should give  special attention   to the following: the
symmetry properties  given by  (\ref{1.13d}) look satisfactory
only at real values of  parameter $a$ -- in this case it describes
symmetry of the  Maxwell equations under Euclidean rotations.
However, if the values of $a$ are imaginary
  the above transformation  $S$  gives  a  Lorentzian boost; for instance, in the plane $0-3$ the boost is
  \begin{eqnarray}
a = i b  \; ,  \qquad S(a=ib) = \left | \begin{array}{cccc}
1  &       0   &      0  &  0  \\
0  &  ch \; b   &  -i sh\; b   &  0  \\
0  & i sh\; b    &  ch\; b  &  0  \\
0 & 0  & 0  &  1
\end{array} \right |
\label{1.15a}
\end{eqnarray}

\noindent and the formulas (\ref{1.13c}) will take the form
\begin{eqnarray}
S\alpha^{1}S^{-1} = \mbox{ch}\; b \; \alpha^{1} + i \mbox{sh}\; b
\; \alpha^{2}
 \nonumber
\\
S\alpha^{2}S^{-1} = -i \mbox{sh}\; b\; \alpha^{1} + \mbox{ch}\; b
\; \alpha^{2} \; , \qquad S\alpha^{3}S^{-1} = \alpha^{3} \;.
\label{1.15b}
\end{eqnarray}

\noindent Correspondingly, the Maxwell matrix equation after
transformation (1.15a,b) will look  asymmetric
\begin{eqnarray}
[\; (-i \partial_{0} + \alpha^{3}  \partial_{3} ) +
 (\mbox{ch}\; b \; \alpha^{1} + i \mbox{sh}\; b  \; \alpha^{2})  \;\partial_{2}
\nonumber
\\
 + \;
(-i \mbox{sh}\; b\; \alpha^{1} + \mbox{ch}\; b \; \alpha^{2} )  \;
\partial_{3} \; ] \;  \Psi'  = SJ \; . \label{1.15c}
\end{eqnarray}

\noindent One  can note  an identity
\begin{eqnarray}
(\mbox{ch}\; b - i \mbox{sh}\; b \; \alpha^{3}) (-i \partial_{0} +
\alpha^{3}
\partial_{3})
\nonumber
\\
= -i (\mbox{ch}\; b \; \partial _{0}  - \mbox{sh}\; b \;
\partial_{3}) + \alpha^{3} ( -sh\; b \; \partial_{0} + \mbox{ch}\;
b \; \partial _{3}) = -i \partial_{0}' + \alpha^{3} \partial_{3}'
\; , \label{1.16a}
\end{eqnarray}

\noindent where derivatives are changed in accordance with the
Lorentzian rule:
\begin{eqnarray}
ch\; b \; \partial _{0}  - sh\; b \; \partial_{3} =
\partial_{0}'  \; , \qquad
 -sh\; b \; \partial_{0} + ch\; b \; \partial _{3} =
\partial_{3}' \; .
\nonumber
\end{eqnarray}

\noindent It remains to determine  the action of the  operator
\begin{eqnarray}
\Delta = \mbox{ch}\; b - i \; \mbox{sh}\; b \; \alpha^{3}
\label{1.16b}
\end{eqnarray}

\noindent on two  other terms in eq. (\ref{1.15c}) -- one might
expect two relations:
\begin{eqnarray}
(\mbox{ch}\; b - i \mbox{sh}\; b \; \alpha^{3}) (\mbox{ch}\; b \;
\alpha^{1} + i \mbox{sh}\; b \; \alpha^{2}) = \alpha^{2}
\nonumber
\\
(\mbox{ch}\; b - i \mbox{sh}\; b \; \alpha^{3}) (-i \mbox{sh}\;
b\; \alpha^{1} + \mbox{ch}\; b \; \alpha^{2} ) = \alpha^{3}  \; .
\label{1.16c}
\end{eqnarray}

\noindent As easily  verified they hold indeed. We should
calculate the term  $\Delta S \; J$:
\begin{eqnarray}
\Delta  S \; J = \left | \begin{array}{c}
\mbox{ch}\; b \; j^{0} + \mbox{sh}\; b \; j^{3} \\
i \;j^{1} \\
i\;j^{2}\\
i( \mbox{sh}\; b \; j^{0} + \mbox{ch}\;b \; j^{3})
\end{array} \right | \; ;
\label{1.16d}
\end{eqnarray}

\noindent it is what  needed. Thus, the symmetry of the matrix
Maxwell equation under the Lorentzian  boost in the plane $0-3$
is described by relations:
\begin{eqnarray}
\Delta (b)\; (-i \partial_{0} + S\alpha^{j}S^{-1}  \partial_{j})\;
\Psi'  = \Delta S \; J \equiv J' \; , \; (-i
\partial_{0}' + \alpha^{j}  \partial_{j}') \Psi'  = J'
\nonumber
\\
S(b) = \left | \begin{array}{cccc}
1  &       0   &      0  &  0  \\
0  &  \mbox{ch} \; b   &  -i \mbox{sh}\; b   &  0  \\
0  & i \mbox{sh}\; b    &  \mbox{ch}\; b  &  0  \\
0 & 0  & 0  &  1
\end{array} \right | \; , \qquad \Delta (b) = \mbox{ch}\; b - i \; \mbox{sh}\; b \; \alpha^{3}\; .
\label{1.17a}
\end{eqnarray}

For the  general case, one can think that for an arbitrary
oriented boost the  operator $\Delta$ should be of the form:
\begin{eqnarray}
\Delta  = \Delta_{\alpha} = ch\; b - i \; sh\; b \; n_{j} \;
\alpha^{j} \; . \nonumber
\end{eqnarray}

To  verify this, one should obtain  mathematical description of
that general  boost. We  will start with the known parametrization
of the real 3-dimension group \cite{1979-Fedorov})
\begin{eqnarray}
O (c)  =  I + 2 \; [ \; c_{0} \; {\bf c}\; ^{\times} + ( {\bf c}\;
^{\times})^{2} \; ]\;  , \qquad  ( {\bf c}\; ^{\times})_{kl}=
-\epsilon _{klj} \;a _{j}  \nonumber
\\
O(c)  =\left | \begin{array}{lll}
 1 -2 (c_{2}^{2} + c_{3}^{2})   &   -2c_{0}c_{3} + 2c_{1}c_{2}    &   +2c_{0}c_{2} + 2c_{1}c_{3}  \\
 +2c_{0}c_{3} + 2c_{1}c_{2}     &  1 -2 (c_{3}^{2} + c_{1}^{2})   &   -2c_{0}c_{1} + 2c_{2}c_{3}   \\
 -2c_{0}c_{2} + 2c_{1}c_{3}     &   +2c_{0}c_{1} + 2c_{2}c_{3}    &  1 -2 (c_{1}^{2} + c_{2}^{2})
 \end{array} \right | \; .
\label{1.18a}
\end{eqnarray}

Transition  to a general boost is achieved by the change
\begin{eqnarray}
c_{0}  \; \Longrightarrow \;  \mbox{ch} \; {b \over 2}  \;  ,
\qquad  c_{j} \; \Longrightarrow \;  i\;  \mbox{sh} \; {b \over 2}
\; \; n_{j} \;, \qquad n_{j}n_{j}=1 \nonumber
\end{eqnarray}

thus we  arrive at
\begin{eqnarray}
O(b, {\bf n}) \nonumber
\\
=\left | \begin{array}{rrr}
 1 -  F  (n_{2}^{2} + n_{3}^{2})        &        -i \mbox{sh}\;b \; n_{3} +F n_{1}n_{2}    &
   i \mbox{sh}\;b \; n_{2} +  F  n_{1}n_{3}  \\
  i \mbox{sh}\;b \; n_{3} + F)  n_{1} n_{2}     &  1 - F (n_{3}^{2} + n_{1}^{2})   &
  -ish\;b \; n_{1} + F  n_{2}n_{3}   \\
 -i \mbox{sh}\;b \; n_{2} + F n_{1}n_{3}     &   i \mbox{sh}\;b \; n_{1} +  F  n_{2}n_{3}    &
 1  - F (n_{1}^{2} + n_{2}^{2})
 \end{array} \right |.
\label{1.20}
\end{eqnarray}

where $F = (1-\mbox{ch}\; b)$. We need to examine relation
\begin{eqnarray}
\Delta (b, {\bf n})\; ( \; -i \partial_{0} + \alpha^{i} O_{ij}(b,
{\bf n}) \partial_{j} \; ) \; \Psi'
  = \Delta(b, {\bf n}) S J  \; .
\nonumber
\end{eqnarray}

After rather  long    calculation we  can indeed  prove the
general statement: the matrix Maxwell equation
\begin{eqnarray}
(-i\;\partial_{0} + \alpha^{i} \partial_{i})\; \Psi = J \nonumber
\end{eqnarray}

\noindent is invariant under an arbitrary Lorentzian boost:
\begin{eqnarray}
\Delta (-i\; \partial_{0} + S\alpha^{i}S^{-1} \partial_{i})\; S
\Psi = \Delta SJ \qquad \Longrightarrow \qquad (\partial_{0}' +
\alpha^{i} \partial_{i}')\; \Psi' = J'  \nonumber
\\
S(ib, {\bf n}) = \left | \begin{array}{cc} 1 & 0 \\ 0 & O(ib, {\bf
n})
\end{array} \right |
\nonumber
\\
t' = \mbox{ch}\; \beta \; t +   \mbox{sh}\; \beta \; {\bf n}\;
{\bf x} \; , \qquad {\bf x}'= +{\bf n}  \; \mbox{sh}\; \beta  \; t
+ {\bf x}  + (\mbox{ch}\; \beta -1)\; {\bf n} \; ({\bf n}  {\bf
x})  \nonumber
\\
\partial_{0}' =  \mbox{ch}\; b  \; \partial_{0}  - \;\mbox{sh}\; b\; ( {\bf n}  \nabla )\; , \qquad
\nabla '  =  -\mbox{sh}\; b \; {\bf n}  \; \partial_{0} + [\nabla
+ (\mbox{ch}\;b -1) {\bf n} ({\bf n} \nabla )  \nonumber
\\
j^{'0} =  \mbox{ch}\; b \; j^{0}  + \mbox{sh}\; b \; ({\bf n} {\bf
j}) \;  , \qquad {\bf j}' =  +\mbox{sh}\; b \;{\bf n} \;  j^{0} +
{\bf j} + (ch\; b-1)\;  {\bf n}\;  (  {\bf n} {\bf j})  \;.
\nonumber
\\
\label{1.30b}
\end{eqnarray}

\noindent Invariance of the  matrix equation under Euclidean
rotations is  achieved in a simpler way:
\begin{eqnarray}
(-i\; \partial_{0} + S\alpha^{i}S^{-1} \partial_{i})\; S \Psi = SJ
\qquad \Longrightarrow \qquad (-i\; \partial_{0}' + \alpha^{i}
\partial_{i}')\; \Psi' = J'
\nonumber
\\
S(a, {\bf n}) = \left | \begin{array}{cc} 1 & 0 \\ 0 & O(a, {\bf
n})
\end{array} \right | \; ,
\; t' = t\; , \qquad {\bf x}'= R(a)  {\bf x}  \nonumber
\\
\partial_{0}' =  \partial_{0} \; , \qquad
\nabla '  =  R(a,-{\bf n}) \nabla \; , \;\; j^{'0} =   j^{0} \;  ,
\qquad {\bf j}' =  R(a, {\bf n})  {\bf j} \; . \label{1.30d}
\end{eqnarray}

\section{On the   Maxwell-Minkowski electrodynamics  in  media}

In agrement with Minkowski approach  \cite{1908-Minkowski}, in
presence of a uniform media we should  introduce  two
electromagnetic tensors
 $F^{ab}$ and  $H^{ab}$ that transform independently under the Lorentz group.
 At this, the known constitutive (or material) relations change their form in the moving reference frame.
In the rest media reference frame the Maxwell equations  are
\begin{eqnarray}
F^{ab} =({\bf E}, \; c{\bf B}), \qquad \mbox{div} \; {\bf B} = 0
\; , \qquad \mbox{rot} \;{\bf E} = -{\partial c{\bf B} \over
\partial c  t}  \nonumber
\\
H^{ab} = ({\bf D}, \; {\bf H}/c)\; , \qquad \mbox{div}\; {\bf D} =
\rho \; , \qquad
 \mbox{rot} \; { {\bf H} \over  c}  = { {\bf J} \over c }  +
  {\partial {\bf D} \over \partial c t} \; .
\label{5.1b}
\end{eqnarray}

\noindent Quantities with simple transformation laws under the
Lorentz group are
\begin{eqnarray}
{\bf f} = {\bf E} + i c {\bf B}   \; , \qquad  {\bf h} =  {1 \over
\epsilon_{0}} \; ({\bf D} + i {\bf H}  / c )\;  , \;\; j^{a} =
(j^{0}=\rho, \; {\bf j} = {\bf J}/c)\; ; \label{5.2}
\end{eqnarray}

\noindent where  ${\bf f}, {\bf h}$ are  complex 3-vector under
complex orthogonal group $SO(3.C)$, the latter
 is isomorphic  to the  Lorentz  group.
One can combine eqs. (\ref{5.1b})  into following ones
\begin{eqnarray}
 \mbox{div}\; ( {{\bf D}\over \epsilon_{0}}  + i \;c
 {\bf B}) = {1 \over \epsilon_{0}}\; \rho
 \nonumber
 \\
\qquad - i \partial_{0}  ( {{\bf D}\over \epsilon_{0}}  + i c {\bf
B}) +\mbox{rot}\; ( {\bf E} + i{{\bf H}/c \over \epsilon_{0}} ) =
\;{i \over \epsilon_{0}} \;{\bf j} \; . \label{5.3}
\end{eqnarray}

\noindent Eqs.  (\ref{5.3})  can be  rewritten in  the form
\begin{eqnarray}
\hspace{40mm} \mbox{div} \; (  { {\bf h } + {\bf h}^{*} \over 2 }
+ { {\bf f} - {\bf f}^{*} \over 2}) = {1 \over \epsilon_{0}}\;
\rho  \nonumber
\\
-i\partial_{0} ( { {\bf h } + {\bf h}^{*} \over 2 } + { {\bf f} -
{\bf f}^{*} \over 2}) + \mbox{rot} \; ({ {\bf f } + {\bf f}^{*}
\over 2 } + { {\bf h} - {\bf h}^{*} \over 2} )  = {i \over
\epsilon_{0}}\; {\bf j} \; . \label{5.5}
\end{eqnarray}

\noindent It has a sense to define two quantities:
\begin{eqnarray}
{\bf M} =   { {\bf h } + {\bf f}  \over 2 } \; , \qquad {\bf N} =
{ {\bf h }^{*} - {\bf f}^{*} \over 2 }  \; , \label{5.6a}
\end{eqnarray}

\noindent which  are  different 3-vectors under the group
$SO(3.C)$: ${\bf M}' = O \;{\bf M} \; , \;  {\bf N}' = O^{*}
\;{\bf N} \;$.
%\label{5.6b}
With respect to Euclidean rotations,  the identity  $O^{*}=O$
holds; whereas for  Lorentzian boosts we have quite  other
identity $O^{*}=O^{-1}$. In terms of ${\bf M}, {\bf N}$, eqs.
(\ref{5.5})  look
\begin{eqnarray}
\mbox{div}\; {\bf M} +  \mbox{div}\; {\bf N} = {1 \over
\epsilon_{0}} \;\rho \; , \qquad -i\partial_{0} {\bf M}  +
\mbox{rot} \; {\bf M} -i\partial_{0} {\bf N}  - \mbox{rot} \;{\bf
N} = {i \over \epsilon_{0}} \;{\bf j} \nonumber
\end{eqnarray}

\noindent or in a matrix form
\begin{eqnarray}
(-i\partial_{0} + \alpha^{i} \partial _{i}) \;  M  + (
-i\partial_{0} + \beta^{i} \partial _{i})  \;  N  = J \nonumber
\\
M = \left | \begin{array}{c} 0 \\ {\bf M} \end{array} \right | \;,
 \qquad N = \left | \begin{array}{c} 0 \\ {\bf N} \end{array} \right | \;, \qquad
J = {1 \over \epsilon_{0}} \; \left | \begin{array}{c} \rho  \\
i\; {\bf j} \end{array} \right | \; . \label{5.8}
\end{eqnarray}

The matrices  $\alpha^{i} $  and  $\beta^{i}$  are taken in the
form
\begin{eqnarray}
\alpha^{1} = \left | \begin{array}{rrrr}
0 & 1  &  0  & 0  \\
-1 & 0  &  0  & 0  \\
0 & 0  &  0  & -1  \\
0 & 0  &  1  & 0
\end{array}  \right |,
\alpha^{2} = \left | \begin{array}{rrrr}
0 & 0  &  1  & 0  \\
0 & 0  &  0  & 1  \\
-1 & 0  &  0  & 0  \\
0 & -1  & 0  & 0
\end{array}  \right |,
\alpha^{3} = \left | \begin{array}{rrrr}
0 & 0  &  0  & 1  \\
0 & 0  & -1  & 0  \\
0 & 1  &  0  & 0  \\
-1 & 0  &  0  & 0
\end{array}  \right |\;\;
\nonumber
\\
\beta^{1} = \left | \begin{array}{rrrr}
0 & 1  &  0  & 0  \\
-1 & 0  &  0  & 0  \\
0 & 0  &  0  & 1  \\
0 & 0  &  -1  & 0
\end{array}  \right |,
\beta^{2} = \left | \begin{array}{rrrr}
0 & 0  &  1  & 0  \\
0 & 0  &  0  & -1  \\
-1 & 0  &  0  & 0  \\
0 & 1  & 0  & 0
\end{array}  \right |,
\beta^{3} = \left | \begin{array}{rrrr}
0 & 0  &  0  & 1  \\
0 & 0  & 1  & 0  \\
0 & -1  &  0  & 0  \\
-1 & 0  &  0  & 0
\end{array}  \right | \; .
\nonumber
\end{eqnarray}

\noindent All of them after squaring give  $-I$, and $\alpha_{i}$
commute with $\beta_{j}$ .

\section{  Minkowski constitutive  relations in a complex 3-vector form
}

Let us  examine how  the  constitutive  relations  for an uniform
media behave under  the Lorentz transformations. One should  start
with these relation in the rest reference frame
\begin{eqnarray}
{\bf D} = \epsilon_{0} \epsilon \; {\bf E} \; , \qquad \; {{\bf H}
\over c} = {1 \over \mu_{0} \mu}  {1 \over c^{2}} \; c {\bf B} = {
\epsilon_{0}\over \mu} \; c{\bf B} \; . \label{6.1}
\end{eqnarray}

\noindent They can be rewritten as
\begin{eqnarray}
{{\bf h} + {\bf h}^{*} \over 2} =   \epsilon  \; {{\bf f} + {\bf
f}^{*} \over 2} \; , \qquad {{\bf h} - {\bf h}^{*} \over 2} =   {1
\over \mu }  \; {{\bf f} - {\bf f}^{*} \over 2} \; , \label{6.2}
\end{eqnarray}

\noindent from whence it follows
\begin{eqnarray}
2{\bf h} =   (\epsilon +{1 \over \mu }) \; {\bf f}  +
  (\epsilon - {1 \over \mu } ) \; {\bf f}^{*}
\; , \;\; 2{\bf h}^{*} =   (\epsilon +{1 \over \mu }) \; {\bf
f}^{*}  +
 (\epsilon - {1 \over \mu } ) \;  {\bf f} \; .
\label{6.3a}
\end{eqnarray}

\noindent This is a complex form of the constitutive relations
(\ref{6.1}). It should be noted that constitutive relations  can
be resolved under  ${\bf f} , \; {\bf f}^{*}$ as well:
\begin{eqnarray}
2{\bf f} =  ({1 \over \epsilon} + \mu ) \; {\bf h} +
 ({1 \over \epsilon } - \mu ) \;  {\bf h}^{*} \; ,
\;\; 2{\bf f}^{*} =  ({1 \over \epsilon} + \mu ) \; {\bf h}^{*} +
 ({1 \over \epsilon } - \mu ) \;  {\bf h} \; ;
\label{6.3b}
\end{eqnarray}

\noindent these are the same constitutive equations (\ref{6.3a})
in other  form. Now let us take into account  the  Lorentz
transformations:
\begin{eqnarray}
{\bf f}' =  O \; {\bf f} \; , \qquad  {\bf f}^{'*} =  O ^{*}\;
{\bf f}^{*} \; , \qquad  {\bf h}' =  O \;{\bf h} \;, \qquad {\bf
h}^{'*} =  O^{*} \;{\bf h}^{*} \; ; \nonumber
\end{eqnarray}

\noindent then eqs. (\ref{6.2}) will become
\begin{eqnarray}
 { O^{-1} {\bf h}' + (O^{-1})^{*}{\bf h}^{*'} \over 2} =
 \epsilon  \; { O^{-1} {\bf f} '+  (O^{-1})^{*} {\bf f}^{*'} \over 2}
\nonumber
\\
{ O^{-1} {\bf h}' -  (O^{-1})^{*}{\bf h}^{*'} \over 2} = {1 \over
\mu }  \; { O^{-1} {\bf f}' - (O^{-1})^{*}{\bf f}^{*'} \over 2} \;
. \nonumber
\end{eqnarray}

\noindent Multiplying both equations by  $O$  and summing  (or
subtracting) the results we get
\begin{eqnarray}
2{\bf h}' =  (\epsilon +{1 \over \mu }) \;{\bf f}' +
 (\epsilon - {1 \over \mu } )\; O(O^{-1})^{*} \;{\bf f}^{'*}
\nonumber
\\
2{\bf h}^{'*} =   (\epsilon +{1 \over \mu }) \;{\bf f}^{' *}+
(\epsilon - {1 \over \mu } ) \;O^{*}O^{-1} \;{\bf f}^{'} \; .
\label{6.5a}
\end{eqnarray}

\noindent Analogously, starting from  (\ref{6.3b}) we can produce
\begin{eqnarray}
2{\bf f}' =  ({1 \over \epsilon} + \mu ) \; {\bf h}' +
 ({1 \over \epsilon } - \mu ) \; O(O^{-1})^{*} \; {\bf h}^{'*}
\nonumber
\\
2{\bf f}^{'*} =  ({1 \over \epsilon} + \mu )  \; {\bf h}^{'*} +
 ({1 \over \epsilon } - \mu ) \;O^{*}O^{-1}  \;{\bf h}' \; .
\label{6.5b}
\end{eqnarray}

Equations   (\ref{6.5a})-(\ref{6.5b}) represent the constitutive
relations  after  changing the  reference frame. In this point one
should distinguish between two cases: Euclidean rotation and
Lorentzian boosts. Indeed, for any  Euclidean rotations
\begin{eqnarray}
O^{*} = O , \qquad \Longrightarrow \qquad O(O^{-1})^{*}  = I,
\qquad  O^{*}O^{-1} = I \; ; \nonumber
\end{eqnarray}

\noindent and therefore   eqs.  (\ref{6.5a})-(\ref{6.5b}) take the
form of  (\ref{6.3a})-(\ref{6.3b}); in other words, at Euclidean
rotations the constitutive  relations do not change their form.
However, for any pseudo-Euclidean rotations (Lorentzian boosts)
\begin{eqnarray}
O^{*} = O^{-1} , \qquad \Longrightarrow \qquad O(O^{-1})^{*}  =
0^{2}, \qquad  O^{*}O^{-1} = O^{*2} \; ; \nonumber
\end{eqnarray}

\noindent  and  eqs. (\ref{6.5a})-(\ref{6.5b})  look
\begin{eqnarray}
2{\bf h}' =   (\epsilon +{1 \over \mu }) {\bf f}' +
 (\epsilon - {1 \over \mu } ) O^{2} \; {\bf f}^{'*}  \; ,
\; 2{\bf h}^{'*} =  (\epsilon +{1 \over \mu }) {\bf f}^{' *}+
 (\epsilon - {1 \over \mu } ) O^{2} {\bf f}^{'}
 %\label{6.6a}
\nonumber
\\
2{\bf f}' =  ({1 \over \epsilon} + \mu )  {\bf h}' +
 ({1 \over \epsilon } - \mu )  O^{*2}  {\bf h}^{'*} \; ,
\; 2{\bf f}^{'*} = ({1 \over \epsilon} + \mu )   {\bf h}^{'*} +
({1 \over \epsilon } - \mu ) O^{*2}  {\bf h}' \; . \nonumber
\\
\label{6.6b}
\end{eqnarray}

In complex 3-vector  form these relations seem to be shorter than
in real 3-vector form:
\begin{eqnarray}
2{\bf D}'  =  \epsilon_{0} \epsilon  \;  [ \;  ( I  + {O O  +
O^{*} O^{*} \over 2} )\; {\bf  E}'\; + {O O  - O^{*} O^{*} \over
2i} \;   c{\bf  B}'\; ]  \nonumber
\\
+  {\epsilon_{0} \over \mu }  \; [ \;( I  - {O O  + O^{*} O^{*}
\over 2} ) \;   {\bf  E}'\; - {OO  - O^{*} O^{*} \over 2i } \;  c
{\bf  B}'\; ]  \nonumber
\\
2{\bf H}' /c =  \epsilon_{0} \epsilon  \;  [ \; (I  - {OO + O^{*}
O^{*} \over 2} ) \;c {\bf  B}'\; + {OO  - O^{*} O^{*} \over 2i} \;
{\bf  E}'\; ]  \nonumber
\\
+ {\epsilon_{0} \over \mu }  \; [ \; (I + {OO  + O^{*} O^{*} \over
2} )\;   c{\bf B}'\; - {O O  - O^{*} O^{*} \over 2i } \;   {\bf
E}'\; ] \; . \label{6.8a}
\end{eqnarray}

\noindent They can be written differently
\begin{eqnarray}
{\bf D}'  =  {\epsilon_{0}  \over 2 } \;  \{ [\; ( \epsilon \; +
{1 \over \mu })  +  ( \epsilon  \; -{1 \over \mu }) \; \mbox{Re}
\; O^{2} \;  ]\;  {\bf  E}'\; + ( \epsilon  \; - {1 \over \mu })
\; \mbox{Im} \; O^{2} \;  c{\bf  B}'\; ] \; \}  \nonumber
\\
 {{\bf H}' \over c } =  {\epsilon_{0} \over 2}   \;  \{
 [
 ( \epsilon  \; + {1 \over \mu })  -  ( \epsilon  \; -{1 \over \mu }) \; \mbox{Re} \; O^{2} \;  ]\;  c{\bf  B}'\;
+ ( \epsilon  \; - {1 \over \mu })   \; \mbox{Im} \; O^{2} \; {\bf
E}'\; ] \;    \} \; . \nonumber
\end{eqnarray}

The matrix $O^{2}$  can be presented differently with the help of
double angle variable:
\begin{eqnarray}
O^{2} =\left | \begin{array}{ccc} \mbox{ch}\;2b  + G \;  n_{1}^{2}
& G \;  n_{1}n_{2}   -  i \; \mbox{sh}\;2 b  \;  n_{3}   &
G \;  n_{3}n_{1}   +  i   \; \mbox{sh}\;2b \; n_{2}  \\[2mm]
G\;  n_{1}n_{2}   +  i   \; \mbox{sh}\;2b \; n_{3} & ch\;2b + G\;
n_{2}^{2}  &
G \;  n_{2}n_{3}  -  i  \; \mbox{sh}\;2b \; n_{1} \\[2mm]
G \;  n_{3}n_{1}   -  i   \; \mbox{sh}\;2b \; n_{2}        & G \;
n_{2}n_{3}   +  i   \; \mbox{sh}\;2b \; n_{1} & \mbox{ch}\;2b +
G\;  n_{3}^{2}
\end{array} \right |,
\nonumber
\end{eqnarray}

where $G = (1 -\mbox{ch}\; 2b )$.

The previous result can be easily extended to more generale
medias, let us restrict  ourselves to linear medias. Indeed,
arbitrary linear media is characterized by the following
constitutive equations:
\begin{eqnarray}
{\bf D}= \epsilon_{0} \;\epsilon(x) \; {\bf E}+ \epsilon_{0}c
\;\alpha(x) \; {\bf B} \; , \qquad {\bf H}= \epsilon_{0}c
\;\beta(x) \; {\bf E}+ {1 \over \mu_{0}} \;\mu(x) \; {\bf B} \; ,
\label{6.16}
\end{eqnarray}

\noindent where $\epsilon(x), \mu(x), \alpha(x), \beta(x)$ are $3
\times 3$  dimensionless matrices. Eqs. (\ref{6.16})  should be
rewritten in terms of complex vectors ${\bf f}, {\bf h}$:
\begin{eqnarray}
{{\bf h} + {\bf h}^{*} \over 2}  = \epsilon(x) \; {{\bf f} + {\bf
f}^{*} \over 2} + \alpha(x) \; {{\bf f} - {\bf f}^{*} \over 2i
}\;\; \nonumber
\\
 {{\bf h} - {\bf h}^{*} \over 2i}  = \beta(x) \;  {{\bf f}
+ {\bf f}^{*} \over 2} + \mu(x) \; {{\bf f} - {\bf f}^{*} \over 2i
}  \; . \label{6.16b}
\end{eqnarray}

\noindent From (\ref{6.16}) it follows
\begin{eqnarray}
{\bf h}=
  [ \; (\epsilon (x) + \mu  (x))  +i (\beta (x) - \alpha(x)) \;  ]  \; {\bf f}
  \nonumber
  \\
   +
 [\;
(\epsilon(x) - \mu(x) )  +i (\beta(x) + \alpha (x) ) \;  ] \; {\bf
f}^{*}   \nonumber
\\
{\bf h}^{*}=
 [ \;
(\epsilon (x) + \mu  (x))  -i (\beta (x)  - \alpha (x)) \;  ] \;
{\bf f}^{*} \nonumber
\\
+ [ \; (\epsilon (x)  - \mu  (x) )  -i (\beta  (x) + \alpha (x) )
\;  ]  \; {\bf f} \;  . \label{6.17}
\end{eqnarray}

Under Lorentz transformations,  relations (6.17) will take the
form
\begin{eqnarray}
 {\bf h} ' =
\epsilon_{0} \left [ (\epsilon  (x)+ \mu (x) )  +i (\beta  (x) -
\alpha (x) ) \right ]  {\bf f}'  \nonumber
\\
 + \left [ \;
(\epsilon (x) - \mu  (x))  +i (\beta (x) + \alpha (x) ) \right ]
[O(O^{-1})^{*}] \; {\bf f}^{'*}  \nonumber
\\
{\bf h}^{'*}= \epsilon_{0} \left [ \; (\epsilon  (x) + \mu (x) )
-i (\beta (x) - \alpha (x)) \right ]   {\bf f}^{'*}  \nonumber
\\
 +
\left [ (\epsilon (x) - \mu (x) )  -i (\beta  (x) + \alpha (x))
\right ]  [O^{*}(O^{-1}) ] {\bf f}' . \label{6.19}
\end{eqnarray}

\noindent For Euclidean rotation, the constitutive relations
preserve  their form. For Lorentz boosts we have
\begin{eqnarray}
{\bf h} ' =
  [ \;
(\epsilon  (x)+ \mu (x) )  +i\; (\beta  (x) - \alpha (x)) \;
 ]  {\bf f}'
 \nonumber
 \\
 +  [\; (\epsilon (x) - \mu (x) )
+ i \;(\beta  (x) + \alpha (x) ) \;  ]  O^{2}  {\bf f}^{'*}
\nonumber
\\
{\bf h}^{'*}=
  [ \;
(\epsilon (x) + \mu (x) )  -i \;(\beta (x) - \alpha (x)) \; ]
{\bf f}^{'*} \nonumber
\\
 +  [\; (\epsilon (x) - \mu (x))
-i \; (\beta  (x)+ \alpha (x))\;   ]  O^{*2} {\bf f}' . \nonumber
\\
\label{6.20}
\end{eqnarray}

\noindent They  are the constitutive  equations for arbitrary
linear medias in a moving reference frame (similar formulas were
produced in quaternion formalism in \cite{1985-Berezin},
\cite{1992-Berezin}).

\section{Symmetry  the matrix  Maxwell equation  in a uniform  media }

As noted, Maxwell equations  in any media can be presented  in the
matrix form:
\begin{eqnarray}
(-i\partial_{0} + \alpha^{i} \partial _{i}) \;M  + (
-i\partial_{0} + \beta^{i} \partial _{i}) \; N  = J \; .
\label{7.3b}
\end{eqnarray}

\noindent We are to study symmetry properties of this equation
under complex rotation group SO(3.C). The  terms with
$\alpha^{j}$  matrices  were  examined in Section {\bf 2}),
 the terms with  $\beta^{j}$ matrix are  new.
We restrict ourselves to demonstrating  the Lorentz symmetry of
eq. (\ref{7.3b}) under two simplest transformations.

First, let  us consider  the Euclidean rotation in the plane
$(1-2)$,  we  examine additionally only the term with
$\beta$-matrices:
\begin{eqnarray}
S\beta^{1}S^{-1} =
 \cos a \; \beta^{1} - \sin a  \; \beta^{2} = \beta^{j} O_{j1}
\nonumber
\\
\beta^{2}S^{-1} =
 \sin a\; \beta^{1} + \cos a \; \beta^{2} = \beta^{j} O_{j2}
\nonumber
\\
S \beta^{3}S^{-1} =  \beta^{3}   = \beta^{j} O_{j3} \; .
\label{7.5c}
\end{eqnarray}

\noindent Therefore, we conclude that eq.  (\ref{7.3b})  is
symmetrical under Euclidean rotations in accordance with the
relations
\begin{eqnarray}
(-i\partial_{0} + S \alpha^{i}S^{-1} \partial _{i})\; M'  +
(-i\partial_{0} + S \beta^{i} S^{-1} \partial _{i}) \; N' = + S J
\; , \;\;\; \Longrightarrow \nonumber
\\
(-i\partial_{0} +  \alpha^{i}  \partial' _{i}) \; M' +
(-i\partial_{0} +  \beta^{i}  \partial ' _{i}) \; N'   = +  J' \;
. \; \label{7.6}
\end{eqnarray}

For the Lorentz boost in the plane $(0-3)$ we have
\begin{eqnarray}
M' = S M \; , \qquad N' = S^{*} N = S^{-1} N , \qquad  S^{*} =
S^{-1} \; ; \nonumber
\end{eqnarray}

\noindent and eq.  (\ref{7.3b}) takes the form (note that the
additional transformation
 $\Delta = \Delta_{(\alpha)}$  is combined  in terms of  $\alpha^{j}$; see Sec. {\bf 2})
\begin{eqnarray}
\Delta_{(\alpha )}  S \; \left [ \; (-i\partial_{0} + \alpha^{i}
\partial _{i})\; S^{-1} M'   + ( -i\partial_{0} + \beta^{i}
\partial _{i})\;S  N' \;  \right ]  =  \Delta S J
\nonumber
\end{eqnarray}

\noindent or
\begin{eqnarray}
\Delta _{(\alpha )} \; \left [ (-i\partial_{0} + S\alpha^{i}
S^{-1}  \partial _{i})\;  M'   + S^{2} ( -i\partial_{0}   + S^{-1}
\beta^{i} S\partial _{i})\; N' \right ]  = J' \; , \nonumber
\end{eqnarray}

\noindent and further
\begin{eqnarray}
(-i\partial_{0}' + \alpha^{i}  \partial' _{i})\;  M'   +
\Delta_{(\alpha )} S^{2} ( -i\partial_{0}   +  S^{-1} \beta^{i}
S\partial _{i})\; N'   = J' \; . \label{7.7c}
\end{eqnarray}

\noindent It remains to prove the relationship
\begin{eqnarray}
\Delta _{(\alpha )} S^{2} \; ( -i\partial_{0}   +  S^{-1}
\beta^{i} S\partial _{i})\; N' = ( -i\partial_{0}'   +   \beta^{i}
\partial _{i} ')\; N' \; .
\label{7.8}
\end{eqnarray}

\noindent  By simplicity reason one may expect two  identities:
\begin{eqnarray}
\Delta _{(\alpha )} S^{2}= \Delta _{(\beta )}  \qquad
\Longleftrightarrow \qquad \Delta _{(\alpha )} S = \Delta _{(\beta
)} S^{-1} \; ,
 \label{7.9a}
\end{eqnarray}
and
\begin{eqnarray}
\Delta _{(\beta )} ( -i\partial_{0}   +  S^{-1} \beta^{i}
S\partial _{i})\; N' = ( -i\partial_{0} '  +   \beta^{i} \partial
_{i}')\; N'  \; . \label{7.9b}
\end{eqnarray}

Let us prove them for a Lorentzian boost in the plane $0-3$:
\begin{eqnarray}
S= \left | \begin{array}{cccc}
1  &       0   &      0  &  0  \\
0  &  ch \; b   &  -i \; sh\; b  &  0  \\
0  & i\; sh \; b   & ch \; b  &  0  \\
0 & 0  & 0  &  1
\end{array} \right | \; , \qquad
S^{-1} = \left | \begin{array}{cccc}
1  &       0   &      0  &  0  \\
0  &  ch \; b   &  -i \; sh\; b  &  0  \\
0  & i\; sh \; b   & ch \; b  &  0  \\
0 & 0  & 0  &  1
\end{array} \right | \; ;
\nonumber
\end{eqnarray}

\noindent we readily  get
\begin{eqnarray}
S^{-1} \beta^{1}S =  ch \; b \; \beta^{1} - i \; sh \; b  \;
\beta^{2} = \beta^{j} O^{-1}_{j1} , \nonumber
\\
S^{-1} \beta^{2}S =
  i \; sh\; b  \beta^{1} + ch\; b  \; \beta^{2} = \beta^{j} O^{-1}_{j2}\; ,
\;\; S^{-1} \beta^{3 }S = \beta^{3} =  \beta^{j} O^{-1}_{j3}\; .
\label{7.10}
\end{eqnarray}

\noindent To verify identity $ \Delta _{(\alpha )} S = \Delta
_{(\beta )} S^{-1} \; , $:
\begin{eqnarray}
(ch\; b - i sh\; b \; \alpha^{3}) S = (ch\; b - i sh\; b \;
\beta^{3}) S^{-1} \; , \nonumber
\end{eqnarray}

\noindent let us calculate  separately the left and right  parts:
\begin{eqnarray}
(ch\; b - i sh\; b \; \alpha^{3}) S =
 (ch\; b - i sh\; b \; \beta^{3}) S^{-1} =
 \left | \begin{array}{cccc}
ch\; b  &   0   &   0  &  -i\; sh\; b  \\
0  &   1   &   0  &  0  \\
0  &   0   &   1  &  0  \\
i\; sh\; b   & 0  & 0  &  ch\; b
\end{array} \right | \; .
\nonumber
\end{eqnarray}

\noindent they coincide with each other, so eq.  (\ref{7.9a})
holds. It remains to prove relation   (\ref{7.9b}). Allowing for
the properties of $\beta$--matrices
\begin{eqnarray}
(\beta^{0})^{2} = -I , \; (\beta^{1}) ^{2} = -I , \; \beta ^{1}
\beta ^{2} =  - \; \beta^{3} ,  \qquad \beta ^{2} \beta ^{1} =  +
\; \beta^{3}  \;\; \mbox{and so on} \nonumber
\end{eqnarray}

\noindent we readily  find
\begin{eqnarray}
\Delta _{(\beta )} \; ( -i\partial_{0}   +  S^{-1} \beta^{i}
S\partial _{i})\; N' =
 (ch\; b - i sh\; b \; \beta^{3}) \; [\; -i \partial_{0} + \beta^{3} \partial_{3}
\nonumber
\\
+   (ch \; b \; \beta^{1} - i \; sh \; b  \; \beta^{2})  \;
\partial_{1} +
   ( i \; sh\; b  \beta^{1}  +  ch\; b  \; \beta^{2} ) \; \partial_{2} \; ] \; N'
\nonumber
\\
= [\; -i (ch\;b \; \partial_{0} - sh\; b \; \partial_{3}) +
\beta^{3} \; (-sh\; b \; \partial_{0} +
 ch\; b \; \partial_{3}) + \beta^{1} \; \partial_{1} + \beta^{2}\;  \partial_{2} \;] \; N' \; ,
\nonumber
\end{eqnarray}

\noindent that is
\begin{eqnarray}
\Delta _{(\beta )} ( -i\partial_{0}   +  S^{-1} \beta^{i}
S\partial _{i})\; N' = ( -i\partial_{0}'    +   \beta^{1} \partial
_{1} + \beta^{2} \partial _{2} + \beta^{3} \partial' _{3} )\; N'
\; ; \label{7.11b}
\end{eqnarray}

\noindent the  relation  (\ref{7.9b}) holds. Thus, the symmetry of
the matrix Maxwell equation in media  under the Lorentz group  is
proved.

\section{Maxwell theory, Dirac matrices and electromagnetic 4-vectors}

Let us  shortly discuss two points relevant to the above  matrix
formulation of the Maxwell theory.

First, let us  write down  explicit form for Dirac matrices in
spinor basis:
\begin{eqnarray}
\gamma^{0} = \left | \begin{array}{cccc}
0 & 0 & 1 & 0 \\
0 & 0 & 0 & 1 \\
1 & 0 & 0 & 0 \\
0 & 1 & 0 & 0
\end{array} \right | , \qquad \gamma^{5} = -i \gamma^{0} \gamma^{1} \gamma^{2} \gamma^{3} =
\left | \begin{array}{cccc}
-1 & 0 & 0 & 0 \\
0 & -1 & 0 & 0 \\
0 & 0 & 1 & 0 \\
1 & 0 & 0 & 1
\end{array} \right | \;
\nonumber
\\
\gamma^{1} = \left | \begin{array}{cccc}
0 & 0 & 0 & -1 \\
0 & 0 & -1 & 0 \\
0 & 1 & 0 & 0 \\
1 & 0 & 0 & 0
\end{array} \right | ,
\gamma^{2} = \left | \begin{array}{cccc}
0 & 0 & 0 & i \\
0 & 0 & -i & 0 \\
0 & -i & 0 & 0 \\
i & 0 & 0 & 0
\end{array} \right | ,
\gamma^{3} = \left | \begin{array}{cccc}
0 & 0 & -1 & 0 \\
0 & 0 & 0 & 1 \\
1 & 0 & 0 & 0 \\
0 & -1 & 0 & 0
\end{array} \right | .
\nonumber
\end{eqnarray}

Taking in mind expressions  for $\alpha^{i}, \beta^{i}$, we
immediately  see the identities
\begin{eqnarray}
\alpha^{1}  =i \gamma^{0} \gamma^{2} , \qquad \alpha^{2}  =
\gamma^{0} \gamma^{5} , \qquad \alpha^{3}  = i \gamma^{5}
\gamma^{2}  \; \nonumber
\\
\beta^{1} = - \gamma^{3} \gamma^{1} , \qquad \beta^{2} = -
\gamma^{3} , \qquad \beta^{3} = - \gamma^{1}  \; ,\label{a2}
\end{eqnarray}

\noindent so the Maxwell matrix equation in media takes  the form
\begin{eqnarray}
(-i \partial_{0}  + i \gamma^{0} \gamma^{2} \partial_{1} +
\gamma^{0} \gamma^{5} \partial_{2}+ i \gamma^{5} \gamma^{2}
\partial_{3} ) \; M  \; \nonumber
\\
+ \; ( -i \partial_{0} - \gamma^{3} \gamma^{1} \partial_{1} -
\gamma^{3} \partial_{2} - \gamma^{1} \partial_{3} ) \; N = J\; .
\label{a3}
\end{eqnarray}

\noindent This Dirac matrix-based form does not seem to be  very
useful to apply in the  Maxwell theory,  it does not prove much
similarity with ordinary Dirac equation (though  that  analogy was
often discussed in the literature).

Now starting from
 electromagnetic 2-tensor and dual to it:
\begin{eqnarray}
\tilde{F}_{\rho\sigma}= {1 \over 2} \; \epsilon_{\rho\sigma\alpha
\beta} F^{\alpha \beta} \; , \qquad F_{\alpha\beta}=  -{1 \over 2}
\; \epsilon_{\alpha \beta \rho \sigma} \tilde{F} ^{\rho \sigma}
\nonumber
\end{eqnarray}

\noindent let us introduce two  electromagnetic 4-vectors (below
$u^{\alpha}$ is any 4-vector that in general may not coincide with
4-velocity)
\begin{eqnarray}
e^{\alpha} = u_{\beta} F^{\alpha \beta} \; , \qquad b^{\alpha} =
u_{\beta} \tilde{F}^{\alpha \beta} \;  , \qquad u^{\alpha}
u_{\alpha} = 1 \; ; \label{a4}
\end{eqnarray}

\noindent inverse formulas are
\begin{eqnarray}
F^{\alpha \beta}  =  (e^{\alpha} \; u^{\beta} - e^{\beta} \;
u^{\alpha}) - \epsilon^{\alpha \beta \rho \sigma} \; b_{\rho} \;
u_{\sigma} \; \nonumber
\\
 \tilde{F}^{\alpha \beta}  = ( b^{\alpha} \;
u^{\beta} - b^{\beta} \; u^{\alpha} ) + \epsilon^{\alpha \beta
\rho \sigma} \; e_{\rho} \;  u_{\sigma} \; . \label{a5}
\end{eqnarray}

\noindent Such electromagnetic 4-vector are presented always  in
the literature   on the electrodynamics of moving bodies, from the
very beginning of relativistic tensor form of electrodynamics --
see Minkowski \cite{1908-Minkowski}, Gordon \cite{1923-Gordon},
Mandel'stam -- Tamm \cite{1925-Mandel'stam}, \cite{1925-Tamm(1)},
\cite{1925-Tamm(2)}; for instance see Y\'epez --  Brito -- Vargas
\cite{1988-Nunez}. The interest to these  field variables  gets
renewed after Esposito paper \cite{1998-Esposito} in 1998.

In 3-dimensional notation
\begin{eqnarray}
E^{1} = -E_{1} = F^{10} \;  ,  \qquad cB^{1} =cB_{1} =
\tilde{F}^{10} = -F_{23} , \qquad \mbox{and so on} \nonumber
\end{eqnarray}

\noindent the formulas  (\ref{a4})  take the form
\begin{eqnarray}
e^{0} = {\bf u} \; {\bf E} \;  , \qquad {\bf e} = u^{0}\;  {\bf E}
+ c\; {\bf u} \times {\bf B}  \nonumber
\\
b^{0} = c\; {\bf u} \; {\bf B} \;  , \qquad {\bf b} = c\; u^{0}\;
{\bf B} - {\bf u} \times {\bf E} \; , \label{a6}
\end{eqnarray}

\noindent or symbolically
 $(e, b) = U(u) \; ({\bf E}, {\bf B}) $;
and inverse the formulas  (\ref{a5})  look
\begin{eqnarray}
{\bf E} = {\bf e} \; u^{0} - e^{0} \;  {\bf u}   + {\bf b} \times
{\bf u}  \nonumber
\\
c\; {\bf B} = {\bf b} \; u^{0} - b^{0} \;  {\bf u}   - {\bf e}
\times {\bf u} \; . \label{a7}
\end{eqnarray}

\noindent or  in symbolical form $ ({\bf E}, {\bf B}) = U^{-1}(u)
\;  (e, b).$

The above  possibility is often used to produce a special form of
the Maxwell equations. For simplicity,  let us consider the vacuum
case:
\begin{eqnarray}
\partial_{\alpha} F_{\beta \gamma} + \partial_{\beta} F_{\gamma \alpha} +
\partial_{\gamma} F_{\alpha \beta} = 0\; , \qquad
\partial_{\alpha}F^{\alpha \beta} = \epsilon_{0}^{-1} j^{\beta}
\nonumber
\end{eqnarray}

\noindent or differently with the help of the dual tensor:
\begin{eqnarray}
\partial_{\beta}\tilde{F}^{\beta \alpha} = 0
\; , \qquad
\partial_{\alpha}F^{\alpha \beta } = \epsilon_{0}^{-1} j^{\beta}\;
\label{a8}
\end{eqnarray}

\noindent These  can  be transformed to variables $b^{\alpha},
b^{\alpha}$:
\begin{eqnarray}
\partial_{\alpha} ( b^{\alpha} \; u^{\beta} - b^{\beta} \; u^{\alpha}  +
\epsilon^{\alpha \beta \rho \sigma} \; e_{\rho} \;  u_{\sigma} ) =
0  \nonumber
\\
\partial_{\alpha} ( e^{\alpha} \; u^{\beta} - e^{\beta} \;  u^{\alpha} -
\epsilon^{\alpha \beta \rho \sigma} \; b_{\rho} \; u_{\sigma} ) =
\epsilon_{0}^{-1} j^{\beta}\; . \label{a8}
\end{eqnarray}

\noindent They can be combined into equations for complex field
function
\begin{eqnarray}
\Phi^{\alpha} = e^{\alpha}+ ib^{\alpha}\; , \qquad
\partial_{\alpha} \; [ \; \Phi^{\alpha}  u^{\beta} -   \Phi^{\beta} u^{\alpha} +
i \epsilon^{\alpha \beta \rho \sigma} \Phi _{\rho} u_{\sigma}\; ]=
\epsilon_{0}^{-1} j^{\beta} \nonumber
\end{eqnarray}

\noindent or differently
\begin{eqnarray}
\partial_{\alpha} \; [ \;  \delta^{\alpha}_{\gamma} u^{\beta}   -
  \delta^{\beta}_{\gamma} u^{\alpha}   +
i \epsilon^{\alpha \beta \rho \sigma}  g_{\rho \gamma}
u_{\sigma}\; ]\; \Phi^{\gamma} = \epsilon_{0}^{-1} j^{\beta} \; .
\label{a10}
\end{eqnarray}

\noindent This is Esposito's representation \cite{1998-Esposito}
of the Maxwell equations. One may introduce four matrices,
functions of 4-vector $u$:
\begin{eqnarray}
(\Gamma^{\alpha})^{\beta}_{\;\;\gamma}  = \delta^{\alpha}_{\gamma}
u^{\beta}  -
  \delta^{\beta}_{\gamma} u^{\alpha}   +
i \epsilon^{\alpha \beta \rho \sigma}  g_{\rho \gamma}
u_{\sigma}\; , \label{a11}
\end{eqnarray}

\noindent then eq. (\ref{a10}) becomes
\begin{eqnarray}
\partial_{\alpha} (\Gamma^{\alpha})^{\beta}_{\;\;\gamma} \; \Phi^{\gamma} =
\epsilon_{0}^{-1} j^{\beta} \; , \qquad  \mbox{or}\qquad
\Gamma^{\alpha}\partial_{\alpha}  \Phi =  \epsilon_{0}^{-1} j  \;
. \label{a12}
\end{eqnarray}

In the 'rest reference frame'  when $u^{\alpha}= (1, 0,0,0)$, the
matrices $\Gamma^{\alpha}$ become simpler and $\Phi =\Psi$:
\begin{eqnarray}
\Gamma^{0}  =    \left | \begin{array}{cccc}
  0 & 0 & 0 & 0  \\
  0 & -1 &0 & 0  \\
  0 & 0 & -1& 0  \\
  0 & 0 & 0& -1
\end{array} \right |
     ,
\Gamma^{1}=
 \left | \begin{array}{cccc}
  0 & 1 & 0 & 0  \\
  0 & 0 & 0 & 0  \\
  0 & 0 & 0 & i  \\
  0 & 0 & -i& 0
\end{array} \right | \; \;
\nonumber
\\
\Gamma^{2}=
 \left | \begin{array}{cccc}
  0 & 0 & 1 & 0  \\
  0 & 0 & 0 & -i  \\
  0 & 0 & 0 & 0  \\
  0 & i & 0& 0
\end{array} \right |,
\Gamma^{3}=
 \left | \begin{array}{cccc}
  0 & 0 & 0 & 1  \\
  0 & 0 & i &   \\
  0 & -i & 0 & 0  \\
  0 & 0 & 0& 0
\end{array} \right | \; .
\nonumber
\end{eqnarray}

\noindent and eq. (\ref{a12}) takes the form
\begin{eqnarray}
\left | \begin{array}{rrrr}
0  &  \partial_{1}  &    \partial_{2}  &  \partial_{3} \\
0  & -\partial_{0}  &  i\partial_{3}  & -i\partial_{2}  \\
0  & -i\partial_{3}  &   -\partial_{0}  &  i\partial_{1}  \\
0  & i\partial_{2}  &  -i\partial_{1}  & -\partial_{0}
\end{array} \right |
\left | \begin{array}{c} 0 \\ E^{1} + icB^{1} \\ E^{2} + icB^{2}
\\ E^{3} + icB^{3}
\end{array} \right | =
\epsilon^{-1}_{0} \left | \begin{array}{c} \rho  \\ j^{1}  \\
j^{2}  \\ j^{3}
\end{array} \right | = \epsilon^{-1}_{0}
\; j \; , \label{a14}
\end{eqnarray}

\noindent or
\begin{eqnarray}
\mbox{div} \; ({\bf E} + ic{\bf B})  = \epsilon^{-1}_{0} \; \rho\;
, \nonumber
\\
-\partial_{0}  ({\bf E} + ic{\bf B}) - i \; \mbox{rot} \; ({\bf E}
+ ic{\bf B}) = \epsilon^{-1}_{0}
  {\bf j} \; .
\nonumber
\end{eqnarray}

\noindent From whence we get equations
\begin{eqnarray}
 \mbox{div} \; c{\bf B} = 0 \; , \qquad \mbox{rot}
\;{\bf E} = -{\partial c {\bf B} \over \partial ct} \nonumber
\\
 \mbox{div}\; {\bf E} = {\rho \over  \epsilon_{0}} , \qquad
 \mbox{rot} \; c{\bf B} =  {{\bf j} \over \epsilon_{0}}  +
   {\partial {\bf E} \over \partial ct} \; ,
\nonumber
\end{eqnarray}

\noindent which coincides with eqs. (\ref{1.2a}).

\noindent Relations (\ref{a14}) correspond to a special choice of
$\alpha$-matrices:
\begin{eqnarray}
\beta  \; (-i \alpha^{0})  = \Gamma^{0} \; ,\;\; \beta \;
\alpha^{j} = \Gamma^{j}\;, \qquad  \mbox{where}\qquad  \beta =
\left | \begin{array}{rrrr}
1 & 0  &  0  & 0  \\
0 & -i  & 0  & 0  \\
0 & 0  &  -i  & 0  \\
0 & 0  &  0  & -i
\end{array}  \right | \;  .
\label{a17}
\end{eqnarray}

Esposito's representation of the Maxwell equation at any 4-vector
$u^{\alpha}$ can be easily related to the matrix equation of
Riemann -- Silberstein -- Majorana -- Oppenheimer:
\begin{eqnarray}
(-i \alpha^{0} \partial_{0} + \alpha^{j}\partial_{j}  )   \Psi   =
J \;  , \label{a19a}
\end{eqnarray}

indeed
\begin{eqnarray}
(-i \alpha^{0} \partial_{0} + \alpha^{j}\partial_{j}  )   U^{-1}
\; ( U \Psi  ) = J  \nonumber
\\
-i\alpha^{0} \; U^{-1} = \beta \;\Gamma^{0} \;, \qquad
\alpha^{j}\; U^{-1} = \beta \;\Gamma^{j} \; , \qquad
 U \Psi  = \Phi
\nonumber
\\
\beta \; ( \Gamma^{0} \partial_{0}  + \Gamma^{j} \partial_{j} ) \;
\Phi = J \;  , \qquad \beta^{-1}J = \epsilon_{0}^{-1} (j^{a})
\nonumber
\\[2mm]
( \Gamma^{0} \partial_{0}  + \Gamma^{j} \partial_{j} )  \Phi =
\epsilon_{0}^{-1} \; j  \; . \label{a19b}
\end{eqnarray}

\noindent Eq.  (\ref{a19b})   is  a  matrix representation of the
Maxwell equations in Esposito's  form
 \begin{eqnarray}
\partial_{\alpha} \; [ \;  \delta^{\alpha}_{\gamma} u^{\beta}   -
  \delta^{\beta}_{\gamma} u^{\alpha}   +
i \epsilon^{\alpha \beta \rho \sigma}  g_{\rho \gamma}
u_{\sigma}\; ]\; \Phi^{\gamma} = \epsilon_{0}^{-1} j^{\beta}
\label{a20}
\end{eqnarray}

Evidently,  eqs. (\ref{a19a}) and (\ref{a20}) are equivalent to
each other. There is no ground to consider the form (\ref{a20})
obtained through the trivial use of identity $I = U^{-1}(u) U(u)$
as having certain  especially profound sense. Our point of view
contrasts with the claim by Ivezi\'c
\cite{2001-Ivezic(1)}-\cite{2002-Ivezic}-\cite{2002-Ivezic(2)}-\cite{2002-Ivezic(3)}-\cite{2003-Ivezic}-
\cite{2005-Ivezic(1)}-\cite{2005-Ivezic(2)}-\cite{2005-Ivezic(3)}-\cite{2006-Ivezic}
that eq. (\ref{a20}) has a status of a true Maxwell equation in  a
moving reference frame (at this $u^{\alpha}$ is identified with
4-velocity).

\section{ Maxwell equation in a curved space-time, no media case}

Now the main question is how the above Maxwell  matrix equation
(first consider the  no-media case)
\begin{eqnarray}
(\alpha^{0}
\partial_{0} + \alpha^{j} \partial_{j} )\;\Psi =J \; ,  \qquad
\alpha^{0} = -i I \nonumber
\\
\Psi = \left | \begin{array}{c} 0 \\
{\bf E} + i c{\bf B}
\end{array} \right |  \; , \qquad  J
= {1 \over \epsilon_{0}} \; \left | \begin{array}{c} \rho  \\
i{\bf j}
\end{array} \right |
%\label{8.1a}
\nonumber
\end{eqnarray}

\noindent can be generalized to the case of a curved space-time
background.
 We should expect  existence of an extended equation in the frame of general Tetrode-Weyl-Fock-Ivanenko tetrad
 approach \cite{1928-Tetrode},  \cite{1929-Weyl(1)}, \cite{1928-Ivanenko-Landau}, \cite{1929-Fock(3)}. Such an equation might be of  the following form
\begin{eqnarray}
 \alpha^{\rho}(x)\;[ \;  \partial_{\rho } + A_{\rho}(x)\;  ]
\; \Psi (x)  = J(x)  \nonumber
\\
\alpha^{\rho}(x) = \alpha^{c} \; e_{(c)}^{\rho}(x) \; , \qquad
A_{\rho}(x) = {1 \over 2} j^{ab} \; e_{(a)} ^{\beta} \;
\nabla_{\rho}  e_{(n) \beta} \; . \label{8.1b}
\end{eqnarray}

\noindent $j^{ab}$ stands for generators of 3-vector  field under
complex orthogonal group $SO(3.C)$, their  explicit form will be
given later. Tetrad represents  four covariant vectors  related to
metric tensor by means of a bilinear function $g_{\alpha \beta}
(x) = \eta^{ab} e_{(a)\alpha} e_{(b)\beta} $,  so that all tetrads
referred by  local Lorentz transformations correspond to the same
metric $g_{\alpha \beta}(x)$: $ e_{(a)\alpha}' (x) = L_{a}^{\;\;b}
(x) \;   e_{(b)\alpha} (x)\; . $ Eq.  (\ref{8.1b}) can be
rewritten as
\begin{eqnarray}
\alpha^{c} \; ( \; e_{(c)}^{\rho} \partial_{\rho}  + {1 \over 2}
j^{ab} \gamma_{abc} \; ) \; \Psi = J(x)\; , \label{8.3a}
\end{eqnarray}

\noindent  where Ricci rotation coefficients are used: $
\gamma_{bac} = - \gamma_{abc} = - e_{(b)\beta ;\alpha}
e^{\beta}_{(a)} e^{\alpha}_{(c)} \; . $
 With regard to eq. (\ref{8.1b}),  one should expect symmetry
properties of the  equation under local gauge transformations:
\begin{eqnarray}
 \Psi' (x) =  S(x) \Psi (x)\; ,  \qquad e_{(a)\alpha}' (x)  =  L_{a}^{\;\;b} (x) \;   e_{(b)\alpha} (x)
\nonumber
\\
 \alpha^{\rho}(x)\;[  \partial_{\rho } + A_{\rho}(x) ]   \; \Psi (x)  = J(x) \; ,\;\; \Longrightarrow
 \nonumber
 \\
 \alpha^{'\rho}(x)\;[  \partial_{\rho } + A'_{\rho}(x) ]   \; \Psi' (x)  = J'(x) \; .
\label{8.4}
\end{eqnarray}

\noindent
 We  should consider separately Euclidean  and Lorentzian tetrad  rotations.
In the case of Euclidean rotations we may expect the following
symmetry:

\vspace{2mm} $ \underline{ S= S [a (x), {\bf n} (x)]} $
\begin{eqnarray}
\Psi' = S  \Psi \; , \qquad  \Psi = S ^{-1} \Psi ' , \qquad  S
J(x)  =J'  \nonumber
\\
S\alpha^{\rho}S^{-1} \; (  \partial _{\rho} +  SA_{\rho}S^{-1} + S
\partial_{\rho }S^{-1}  )   \; \Psi '(x)  =  S J(x)
\nonumber
\\
S\alpha^{\rho}S^{-1} = \alpha^{'\rho} \; , \qquad SA_{\rho}S^{-1}
+  S \partial_{\rho }S^{-1}   =  A_{\rho} '\; . \label{8.5a}
\end{eqnarray}

In the case of Lorentzian  rotations we may expect other symmetry
realized in accordance with relations

\vspace{2mm}

$ \underline{ S= S [i b (x) ,  {\bf n} (x)] \; , \;\; \Delta =
\Delta [i b (x), {\bf n} (x)]} $
\begin{eqnarray}
\Psi' = S  \Psi \; , \qquad  \Psi = S ^{-1} \Psi ' , \qquad \Delta
\;  S J(x)  =J'  \nonumber
\\
\Delta S\alpha^{\rho}S^{-1} \; (  \partial _{\alpha} +
SA_{\alpha}S^{-1} +  S \partial_{\alpha }S^{-1}  )   \; \Psi '(x)
=  \Delta S J(x)   \nonumber
\\
\Delta \;  S\alpha^{\rho}S^{-1} = \alpha^{'\rho} \; , \qquad
  SA_{\alpha}S^{-1} +  S \partial_{\alpha }S^{-1}     =  A_{\alpha}' \; .
\label{8.5b}
\end{eqnarray}

Symmetry properties of the local matrices $\alpha^{\rho}(x)$ can
be found quite straightforwardly on the base of analysis performed
for the flat Minkowski  space. Indeed, for  local Euclidean
rotations,  the rule for $S\alpha^{\rho}(x) S^{-1} $ is
\begin{eqnarray}
S\alpha^{\rho}S^{-1} = S \alpha^{0} e_{(0)}^{\rho} S^{-1} + S
\alpha^{l} e_{(l)}^{\rho} S^{-1} \nonumber
\\
=
 \alpha^{0} e_{(0)}^{\rho} + \alpha^{k}O_{kl} e_{(l)}^{\rho}  =
\alpha^{0} e_{(0)}^{\rho} + \alpha^{k}  e_{(k)}^{'\rho}  =
\alpha^{'\rho} \; .
%\label{8.6a}
\nonumber
\end{eqnarray}

\noindent For  local Lorentzian rotations, we can easily prove a
symmetry relation:
\begin{eqnarray}
\Delta \; S\alpha^{\rho} (x) S^{-1} = \Delta \; S \alpha^{a}
e_{(a)}^{\rho} S^{-1} \nonumber
\\
= [\Delta \; S \alpha^{a}  S^{-1} ] \; e_{(a)}^{\rho}=
 \alpha^{b} L_{b}^{\;\; a} \; e_{(a)}^{\rho} =  \alpha^{b} e_{(b)}^{'\rho} =\alpha^{' \rho}(x) \; .
\nonumber
\end{eqnarray}

\noindent The transformation law  for the  complex  3-vector
connection $A_{\rho}(x)$ will be proved in  Section {\bf 8}.

\section{ On tetrad transformation for  complex 3-vector  connection }

First, let us list six elementary rotations from the local group
$SO(3.C)$:

\begin{eqnarray}
S_{23} = \left | \begin{array}{rrrr}
1  &       0   &      0   &  0  \\
0  &       1   &      0   &  0  \\
0  &       0   & \cos a   & -\sin a   \\
0  &       0   & \sin a  & \cos a    \\
\end{array} \right | , \qquad
S_{01} =\left | \begin{array}{rrrr}
1  &       0   &      0  &  0  \\
0  &   0   &  0   &  0  \\
0  & 0    &  \mbox{ch}\; b  &  -i \mbox{sh}\; b  \\
0 & 0  & +i \mbox{sh}\; b  &  \mbox{ch} \; b
\end{array} \right |
\nonumber
\\
S^{1}= j^{23} =  \left | \begin{array}{cc} 0 & 0 \\ 0 & \tau_{1}
\end{array} \right |,
\qquad N^{2} = j^{01} =  +i \left | \begin{array}{cc} 0 & 0 \\ 0 &
\tau_{1}
\end{array} \right |
\nonumber
\\
S_{31} = \left | \begin{array}{rrrr}
1  &       0   &      0   &  0  \\
0  &       \cos a   &      0   &  \sin a  \\
0  &       0   &   1     &  0   \\
0  &      -\sin a     & 0 &  \cos a     \\
\end{array} \right | , \qquad  S_{02} =
\left | \begin{array}{rrrr}
1  &       0   &      0  &  0  \\
0  &  \mbox{ch} \; b   &  0 & + i \mbox{sh}\; b    \\
0  &  0     &  0  &  0  \\
0 &   -i \mbox{sh}\; b& 0  &  \mbox{ch}\; b
\end{array} \right |
\nonumber
\\
S^{2} = j^{31} =
 \left | \begin{array}{cc}
0 & 0 \\ 0 & \tau_{2}
\end{array} \right | \; , \qquad
N^{2} =  j^{02} = +i \left | \begin{array}{cc} 0 & 0 \\ 0 &
\tau_{2}
\end{array} \right |
\nonumber
\\
S_{12} = \left | \begin{array}{rrrr}
1  &       0   &      0  &  0  \\
0  &  \cos a   & -\sin a  &  0  \\
0  & \sin a   & \cos a  &  0  \\
0 & 0  & 0  &  1
\end{array} \right | , \qquad
\qquad  S_{03} = \left | \begin{array}{rrrr}
1  &       0   &      0  &  0  \\
0  &  \mbox{ch} \; b   &  -i \mbox{sh}\; b   &  0  \\
0  & +i \mbox{sh}\; b    &  \mbox{ch}\; b  &  0  \\
0 & 0  & 0  &  1
\end{array} \right |
\nonumber
\\
S^{3} = j^{12} = \left | \begin{array}{cc} 0 & 0 \\ 0 & \tau_{3}
\end{array} \right |,
\qquad N^{3} = j^{03} = +i
 \left | \begin{array}{cc}
0 & 0 \\ 0 & \tau_{3}
\end{array} \right | \; ;
\nonumber
\end{eqnarray}

\noindent they obey the commutative relations:
\begin{eqnarray}
S^{1}  S^{2}  -S^{2} S^{1}   = S^{3} \; , \qquad N^{1}  N^{2}
-N^{2} N^{1}  = -S^{3} \; , \qquad
 S^{1}N^{2} - N^{2}S^{1}  = +N^{3}\; ;
\nonumber
\end{eqnarray}

\noindent and remaining ones written be cyclic symmetry. Let us
turn to some properties of the  connection $A_{\alpha}(x) $:
\begin{eqnarray}
A_{\alpha}(x) = {1 \over 2} j^{ab} \; e_{(a)} ^{\beta} \;
\nabla_{\alpha}  e_{(b) \beta} \nonumber
\\
=
 S^{1} e_{(2)} ^{\beta} \; \nabla_{\alpha}  e_{(3) \beta} + S^{2}
e_{(3)} ^{\beta} \; \nabla_{\alpha}  e_{(2) \beta} + S^{3} e_{(1)}
^{\beta} \; \nabla_{\alpha}  e_{(2) \beta}  \nonumber
\\
+ N^{1} e_{(0)} ^{\beta} \; \nabla_{\alpha}  e_{(1) \beta} + N^{2}
e_{(0)} ^{\beta} \; \nabla_{\alpha}  e_{(2) \beta} + N^{3} e_{(0)}
^{\beta} \; \nabla_{\alpha}  e_{(3) \beta} ]\;. \label{9.1a}
\end{eqnarray}

\noindent
%It is sufficient to examine a non-trivial block $ 3 \times 3$ in this connection $B_{\alpha}(x) $, this quantity
% can be decomposed in a linear combination of three matrices
% $\tau_{k}$. Indeed,

\noindent Taking in mind the identity  $N_{k} = +iS_{k}$,  and
introducing  new complex variables
\begin{eqnarray}
A_{(1) \alpha } = e_{(2)} ^{\beta} \; \nabla_{\alpha}  e_{(3)
\beta} + i\; e_{(0)} ^{\beta} \; \nabla_{\alpha}  e_{(1) \beta} \;
 \nonumber
\\
A_{(2) \alpha } = e_{(3)} ^{\beta} \; \nabla_{\alpha}  e_{(1)
\beta} + i\; e_{(0)} ^{\beta} \; \nabla_{\alpha}  e_{(2) \beta} \;
 \nonumber
\\
A_{(3) \alpha } = e_{(1)} ^{\beta} \; \nabla_{\alpha}  e_{(2)
\beta} + i\; e_{(0)} ^{\beta} \; \nabla_{\alpha}  e_{(3) \beta} \;
, \nonumber
\end{eqnarray}

\noindent  one can read the above connection as
\begin{eqnarray}
A_{\alpha}(x) =  S^{k}\;  A_{(k) \alpha } \; .
 \label{9.1c}
\end{eqnarray}

\noindent With the use of notation
\begin{eqnarray}
A_{\alpha}(x) = {1 \over 2} j^{ab}  e_{(a)} ^{\beta}
\nabla_{\alpha}  e_{(b) \beta}  = {1 \over 2} j^{ab} A_{(a)(b)
\alpha} \; , \;  A_{(a)(b) \alpha}  = -  A_{(b)(a) \alpha}
\label{9.2a}
\end{eqnarray}

\noindent the above definition for   $A_{(k) \alpha }$  can be
rewritten  differently:
\begin{eqnarray}
A_{(1) \alpha } =  A_{(2)(3) \alpha} + i  A_{(0)(1) \alpha}\;
\nonumber
\\
A_{(2) \alpha } =  A_{(3)(1) \alpha} + i  A_{(0)(2) \alpha}\;
\nonumber
\\ A_{(3) \alpha } =  A_{(1)(2) \alpha} + i  A_{(0)(3)
\alpha} \; . \nonumber
\end{eqnarray}

\noindent In other words, the 3-quantity $A_{(k) \alpha }$  with
respect to 3-index
 $(k)$  is constructed  in terms of "tensor"
\hspace{2mm}  $A_{(a)(b) \alpha} $ by the same rule that used at
constructing 3-dimensional complex vector
  $ -i({\bf E}+i c{\bf B})$ in terms of component of tensor $F_{ab}$.

It is readily verified that such  3-dimensional complex vectors
can be built in terms of a skew-symmetric 2-rank  real tensor
through  a simple and symmetrical algebraic construction:
\begin{eqnarray}
{i \over 2} \; \bar{\sigma}^{a} \sigma^{b} A_{(a)(b)\alpha} =
\sigma^{k}  A_{(k)\alpha} \;  , \qquad {i\over 2} \; \sigma^{a}
\bar{\sigma}^{b} A_{(a)(b)\alpha} = \sigma^{k}  A_{(k)\alpha}^{*}
\;   \; . \label{9.5a}
\end{eqnarray}

\noindent From the above it follows covariant formulas for
$A_{(k)\alpha}$ and  $A_{(k)\alpha}^{*}$:
\begin{eqnarray}
 A_{(k)\alpha}  =
{i \over 4}\;  \mbox{Sp}\; [ \bar{\sigma}^{a} \sigma^{b}
A_{(a)(b)\alpha} ] \; , \qquad
 A_{(k)\alpha} ^{*} = {i\over 4} \mbox{Sp}\; [ \sigma_{k}
\sigma^{a}  \bar{\sigma}^{b} A_{(a)(b)\alpha} ] \; . \label{9.5b}
\end{eqnarray}

Now, starting from relations between any two tetrads by a local
Lorentz transformation:
\begin{eqnarray}
e^{'\alpha}_{(a)} = L_{a}^{\;\;b} e_{(b)}^{\alpha} \; , \qquad
e^{\alpha}_{(a)} = (L^{-1})_{a}^{\;\;\;b} e_{(b)}^{'\alpha} \; ,
\qquad
%\label{9.6}
\nonumber
\end{eqnarray}

\noindent let us derive a rule to transform 3-vector connection
when changing the  tetrad:
\begin{eqnarray}
 A_{(a)(b) \alpha} = e_{(a)} ^{\beta} \; \nabla_{\alpha}  e_{(b) \beta}=
 (L^{-1})_{a}^{\;\;\;m} e_{(m)}^{'\beta} \nabla_{\alpha}
(L^{-1})_{b}^{\;\;\;n} e_{(n)\beta}'  \nonumber
\\
= (L^{-1})_{a}^{\;\;\;m} e_{(m)}^{'\beta} (L^{-1})_{b}^{\;\;\;n}
\;\; \nabla_{\alpha} e_{(n)\beta}'+ (L^{-1})_{a}^{\;\;\;m}
e_{(m)}^{'\beta}  \;\; { \partial (L^{-1})_{b}^{\;\;\;n} \over
\partial x^{\alpha}} \;
 e_{(n)\beta}' \; ,
\nonumber
\end{eqnarray}

\noindent that is
\begin{eqnarray}
 A_{(a)(b) \alpha} =
 (L^{-1})_{a}^{\;\;\;m} (L^{-1})_{b}^{\;\;\;n}  A_{(m)(n)\alpha} '  +
(L^{-1})_{a}^{\;\;\;m}   g_{(m)(n)}   { \partial
(L^{-1})_{b}^{\;\;\;n} \over \partial x^{\alpha}} \; . \label{9.7}
\end{eqnarray}

\noindent Let us act on this relation from the  left by an
operator $ {i\over 4} \; \mbox{Sp}\; [ \sigma_{k} \bar{\sigma}^{a}
\sigma^{b} \;  ... ] \; ; $ it results in
\begin{eqnarray}
A_{(k)\alpha} = {i\over 4} \;\mbox{Sp}\;  [ \sigma_{k}
\bar{\sigma}^{a}  \sigma^{b}
 A_{(a)(b) \alpha}   ] \nonumber
 \\
 =
{i\over 4}\; \mbox{Sp}\;  [ \sigma_{k} \bar{\sigma}^{a} \sigma^{b}
 (L^{-1})_{a}^{\;\;\;m} (L^{-1})_{b}^{\;\;\;n} \;  \; A_{(m)(n)\alpha} '  ] \;
    \nonumber
  \\
  +
{i\over 4}\; \mbox{Sp}\;   [ \sigma_{k} \bar{\sigma}^{a}
\sigma^{b} (L^{-1})_{a}^{\;\;\;m}  \; \; g_{(m)(n)}  \;\;
 { \partial (L^{-1})_{b}^{\;\;\;n} \over \partial x^{\alpha}} \;  ] \;  .
\label{9.8}
\end{eqnarray}

\noindent One may expect eq. (\ref{9.8}) to be equivalent to
\begin{eqnarray}
A_{(k)\alpha} = O^{-1}_{kn}  A_{(n)\alpha}' + {i\over 4}
\mbox{Sp}\;   [ \sigma_{k} \bar{\sigma}^{a}  \sigma^{b}
(L^{-1})_{a}^{\;\;\;m}  \; \; g_{(m)(n)}  \;\;
 { \partial  \over \partial x^{\alpha}}  (L^{-1})_{b}^{\;\;\;n} \;  ] \;  ;
\label{9.9}
\end{eqnarray}

\noindent it is so if an identity holds
\begin{eqnarray}
 {i\over 4} \mbox{Sp}\;   [ \sigma_{k}
\bar{\sigma}^{a}  \sigma^{b}
 (L^{-1})_{a}^{\;\;\;m} (L^{-1})_{b}^{\;\;\;n} \;\; A_{(m)(n)\alpha} '  \; ]  = O^{-1}_{kl}  A_{(l)\alpha}'  \; .
\label{9.10}
\end{eqnarray}

\noindent which is proved by direct calculation (all details are
omitted). Now, we are ready to prove the following relationships:
\begin{eqnarray}
OA_{\rho} O^{-1} +O  \partial _{\rho}  O^{-1}   = A_{\rho}' \; .
\label{9.14a}
\end{eqnarray}

\noindent Taking into account the  linear decomposition
$A_{\alpha} = A_{(k) \alpha} \tau_{k}$,
 eq. (\ref{9.14a}) can be rewritten as
\begin{eqnarray}
\tau^{l} O_{lk} A_{(k) \alpha}  +O \partial_{\alpha} O^{-1} =
\tau^{k} A_{(k)\alpha}' \; . \label{9.14b}
\end{eqnarray}

\noindent Substituting expression for $A_{(k) \alpha}$  through
$A_{(k) \alpha}'$ (see   (\ref{9.9}))
\begin{eqnarray}
A_{(k)\alpha} = O^{-1}_{kn}  A_{(n)\alpha}' + {i\over 4} \;
\mbox{Sp}\; [\;  ( \sigma_{k} \bar{\sigma}^{a} \sigma^{b}
(L^{-1})_{a}^{\;\;\;m}  \; \; g_{(m)(n)}  \;\;
 { \partial  \over \partial x^{\alpha}}  (L^{-1})_{b}^{\;\;\;n} \; ] \;  ;
\nonumber
\end{eqnarray}

\noindent we get
\begin{eqnarray}
\tau^{l} O_{lk}  \{ O^{-1}_{kn}  A_{(n)\alpha}' + {i\over 4}
\mbox{Sp}\;  [\;  \sigma_{k} \bar{\sigma}^{a}  \sigma^{b}
(L^{-1})_{a}^{\;\;\;m}  \; \; g_{(m)(n)}  \;\;
 { \partial  \over \partial x^{\alpha}}  (L^{-1})_{b}^{\;\;\;n} \; ] \;
   \}
   \nonumber
   \\
    +
  \; O \partial_{\alpha} O^{-1} = \tau^{k} A_{(k)\alpha}' \; .
\nonumber
\end{eqnarray}

\noindent From whence we conclude that an identity   must hold:
\begin{eqnarray}
\tau^{l} O_{lk} \; {i\over 4} \mbox{Sp}\;  [ \;  \sigma_{k}
\bar{\sigma}^{a}  \sigma^{b} C_{ab, \alpha}  \; ] \;
 +\;  O \partial_{\alpha} O^{-1} = 0 \; ,
\label{9.16}
\end{eqnarray}

\noindent where
\begin{eqnarray}
C_{ab, \alpha}= (L^{-1})_{a}^{\;\;\;m}  \; \; g_{(m)(n)}  \;\;
 { \partial  \over \partial x^{\alpha}}  (L^{-1})_{b}^{\;\;\;n} \;.
\nonumber
\end{eqnarray}

\noindent The identity (\ref{9.16})  holds indeed which can be
proved  with the use of   simplest transformations -- all details
are omitted. Thus, generally covariant Maxwell matrix equation in
a Riemannian space-time possesses all needed symmetry properties
under local tetrad  transformations and therefore it is correct.

\section{  Maxwell   equation in a curved space-time, in  media}

Now we are to extend the Maxwell matrix equation in media to a
curved space-time background: starting from the equation
\begin{eqnarray}
(-i\partial_{0} + \alpha^{i} \partial _{i}) \;M  + (
-i\partial_{0} + \beta^{i} \partial _{i}) \; N  = J  \nonumber
\\
M' = S M \; , \qquad N' = S^{*} N  \; ; \label{10.1}
\end{eqnarray}

\noindent we may propose the following one
\begin{eqnarray}
\alpha_{\rho}(x) ( i\partial_{\rho} + A_{\rho} ) \;M  +
\beta_{\rho}(x) ( i\partial_{\rho} + B_{\rho} ) \;N  = J \; ,
\label{10.2}
\end{eqnarray}

\noindent where $A(x), B_{\rho}=A^{*}(x)$  stand connections
related to the  fields $M(x)$ and $N(x)$ respectively.
 We  should consider separately Euclidean  and Lorentzian tetrad  rotations.

\vspace{2mm}

In the case of \underline{Euclidean rotations} we may expect the
following  symmetry:
\begin{eqnarray}
 S^{*} = S \; ,  \qquad  S(x) J(x)  = J' (x)
\nonumber
\\
M'(x) = S(x) M (x) \; , \qquad  N'(x) = S (x) N (x)
%\label{10.3a}
\nonumber
\\
S\alpha^{\rho}S^{-1} \; (  \partial _{\rho} +  SA_{\rho}S^{-1} + S
\partial_{\rho }S^{-1}  )   \; M'(x)  \nonumber
\\
+ S\beta^{\rho}S^{-1} \; (  \partial _{\rho} +  SB_{\rho}S^{-1} +
S \partial_{\rho }S^{-1}  )   \; N'(x)
 =  S J(x)
%\label{10.3b}
\nonumber
\\
S\alpha^{\rho} S^{-1} = \alpha^{'\rho} \; , \qquad
S\beta^{\rho}S^{-1} = \beta^{'\rho}  \nonumber
\\
SA_{\rho}S^{-1} +  S \partial_{\rho }S^{-1}   =  A_{\rho} '\;,
\qquad SB_{\rho}S^{-1} +  S \partial_{\rho }S^{-1}   =  B_{\rho}
'\; . \label{10.3c}
\end{eqnarray}

In the case of \underline{Lorentzian  rotations}  we may expect
other    symmetry realized in accordance with relations
\begin{eqnarray}
 S^{*}  = S^{-1} \; , \qquad  \Delta_{\alpha}(x) \; , \qquad \Delta_{\alpha} (x) \;  S (x)\; J(x)  =J'
\nonumber
\\
M(x) ' = S (x)  M (x)  \; , \qquad  N'(x) = S ^{*} (x) N '(x) =
S^{-1}(x) N'(x)  \qquad
%\label{10.4a}
\nonumber
\\
\Delta_{\alpha} S\alpha^{\rho}S^{-1} \; (  \partial _{\alpha} +
SA_{\alpha}S^{-1} +  S \partial_{\alpha }S^{-1}  )   \; M '(x)
\nonumber
\\
+ \Delta_{\alpha} S^{2}  \; S^{-1} \beta^{\rho} S \; (  \partial
_{\alpha} +  S^{-1} B_{\alpha}S +  S^{-1} \partial_{\alpha }S  )
\; N '(x) =  \Delta S J(x)
%\label{10.4b}
\nonumber
\\
\Delta_{\alpha} \;  S\alpha^{\rho}S^{-1} = \alpha^{'\rho} \; ,
\qquad
  SA_{\alpha}S^{-1} +  S \partial_{\alpha }S^{-1}     =  A_{\alpha}'
\nonumber
\\
\Delta_{\alpha} S^{2} \;  S^{-1}\beta^{\rho}S = \beta^{'\rho} \; ,
\qquad
  S^{-1}B_{\alpha}S +  S^{-1} \partial_{\alpha }S     =  B_{\alpha}' \; .
\label{10.4c}
\end{eqnarray}

In addition to calculation performed in Sections {\bf 8,9}, we
need to consider only relations involving
 matrices $\beta^{\rho}$ and connection $B_{\rho}$.
For Euclidean rotation:
\begin{eqnarray}
S\beta^{\rho}S^{-1} = S \beta^{0} e_{(0)}^{\rho} S^{-1} + S
\beta^{l} e_{(l)}^{\rho} S^{-1} \nonumber
\\
= \beta^{0} e_{(0)}^{\rho} + \beta^{k}O_{kl} e_{(l)}^{\rho}  =
\beta^{0} e_{(0)}^{\rho} + \beta^{k}  e_{(k)}^{'\rho}  =
\beta^{'\rho} \; . \nonumber
\end{eqnarray}

\noindent For  local Lorentzian rotations
\begin{eqnarray}
\Delta S^{2} \; S^{-1}\beta^{\rho} (x) S = \Delta S^{2} \; S
^{-1}\beta^{a} e_{(a)}^{\rho} S   \nonumber
\\
=[\Delta S^{2} \; (S^{-1} \alpha^{a}  S )\; ] \; e_{(a)}^{\rho}=
 \alpha^{b} L_{b}^{\;\; a} \; e_{(a)}^{\rho} =  \beta^{b} e_{(b)}^{'\rho} =\beta^{' \rho}(x) \; .
\nonumber
\end{eqnarray}

Transformation laws for two connections
\begin{eqnarray}
SA_{\rho}S^{-1} +  S \partial_{\rho }S^{-1}     =  A_{\rho}' \; ,
\qquad
 S^{-1}B_{\rho}S +  S^{-1} \partial_{\rho }S     =  B_{\rho} \; ,
\nonumber
\end{eqnarray}

\noindent in fact are complex conjugated relations, because of
identities
\begin{eqnarray}
S^{-1} = S^{*} , \qquad S= (S^{*})^{-1}, \qquad (B_{\alpha})^{*} =
A_{\alpha} \; , \nonumber
\end{eqnarray}

\noindent so we need not any additional calculation.

\section{Matrix equation in  explicit component form
 }

Now we are going to derive tensor generally covariant Maxwell
equations  when starting with the matrix form
\begin{eqnarray}
-i  \; ( \; e_{(0)}^{\rho} \partial_{\rho}  +  {1 \over 2} j^{ab}
\gamma_{ab0} \; )\Psi + \alpha^{k} \; ( \; e_{(k)}^{\rho}
\partial_{\rho}  + {1 \over 2} j^{ab} \gamma_{abk} \; )\Psi =
J(x)\; . \label{11.2}
\end{eqnarray}

\noindent Taking in mind
\begin{eqnarray}
{1 \over 2} j^{ab} \gamma_{ab0} = [ s^{1} ( \gamma_{230} +i
\gamma_{010} ) + s^{2} ( \gamma_{310} +i \gamma_{020}) + s^{3} (
\gamma_{120} +i \gamma_{030} )\;  ]  \nonumber
\\
{1 \over 2} j^{ab} \gamma_{abk} = [ s^{1} ( \gamma_{23k} +i
\gamma_{01k} ) + s^{2} ( \gamma_{31k} +i \gamma_{02k}) + s^{3} (
\gamma_{12k} +i \gamma_{03k} )\;  ]
%\label{11.3}
\nonumber
\end{eqnarray}

\noindent and introducing notation
\begin{eqnarray}
e_{(0)}^{\rho} \partial_{\rho} = \partial_{(0)} \;, \qquad
e_{(k)}^{\rho} \partial_{\rho} = \partial_{(k)} \nonumber
\\
( \gamma_{01a}, \gamma_{02a} , \gamma_{03a} ) = {\bf v}_{a} \; ,
\qquad ( \gamma_{23a}, \gamma_{31a} , \gamma_{12a} ) = {\bf p}_{a}
\; , \qquad a =0,1,2,3 \nonumber
\end{eqnarray}

\noindent eq. (\ref{11.2}) can be transformed to the form
\begin{eqnarray}
(\;  \alpha^{k} \;   \partial_{(k)}  + {\bf s} {\bf v}_{0}  +
 \alpha^{k} \; {\bf s} {\bf p}_{k} \; )\; \left | \begin{array}{c}
0 \\ {\bf E} + i c{\bf B}
\end{array} \right |
\nonumber
\\
 -i\;
 ( \; \partial_{(0)}   +
  {\bf s} {\bf p}_{0}  - \alpha^{k} {\bf s} {\bf v}_{k}) \;\left | \begin{array}{c}
0 \\ {\bf E} + i c{\bf B}
\end{array} \right |
 = {1 \over \epsilon_{0}}
 \left | \begin{array}{c}
 \rho  \\  i \;{\bf j}
 \end{array} \right | \; .
\label{11.5}
\end{eqnarray}

\noindent Let us divide  equation (\ref{11.5})  into real and
imaginary parts:
\begin{eqnarray}
(  \alpha^{k}    \partial_{(k)}  + {\bf s} {\bf v}_{0}  +
 \alpha^{k}  {\bf s} {\bf p}_{k}  )  \left | \begin{array}{c}
0 \\ {\bf E}
\end{array} \right | +
 (   \partial_{(0)}   +
  {\bf s} {\bf p}_{0}  - \alpha^{k} {\bf s} {\bf v}_{k}) \left | \begin{array}{c}
0 \\ c{\bf B}
\end{array} \right |
 = {1 \over \epsilon_{0}}
 \left | \begin{array}{c}
 \rho  \\   0
 \end{array} \right |\;
 \nonumber
\\
(  \alpha^{k}   \partial_{(k)}  + {\bf s} {\bf v}_{0}  +
 \alpha^{k}  {\bf s} {\bf p}_{k}  ) \left | \begin{array}{c}
0 \\ c{\bf B}
\end{array} \right |-
 (  \partial_{(0)}   +
  {\bf s} {\bf p}_{0}  - \alpha^{k} {\bf s} {\bf v}_{k}) \left | \begin{array}{c}
0 \\ {\bf E}
\end{array} \right |
 = {1 \over \epsilon_{0}}
 \left | \begin{array}{c}
 0  \\   {\bf j}
 \end{array} \right | \; .
\nonumber
\end{eqnarray}

\noindent From  whence we  produce explicit equations (for
shortness let $c=1$):
\begin{eqnarray}
 \partial_{(k)}  E_{k}  -
[ (p_{23} -p_{32}) E_{1}  +  (p_{31}-p_{13}) E_{2}  +
(p_{12}-p_{21}) E_{3} ]  \nonumber
\\
+ [ (v_{23} -v_{32}) B_{1}  +  (v_{31}-v_{13}) B_{2}  +
(v_{12}-v_{21}) B_{3} = {1 \over \epsilon_{0}} \rho\; ,
\label{11.10a}
\\
 \partial_{(k)}  B_{k} -
[ (p_{23} -p_{32}) B_{1}  +  (p_{31}-p_{13}) B_{2}  +
(p_{12}-p_{21}) B_{3} ]  \nonumber
\\
- [ (v_{23} -v_{32}) E_{1}  +  (v_{31}-v_{13}) E_{2}  +
(v_{12}-v_{21}) E_{(3)} =  0 \; , \label{11.10b}
\end{eqnarray}
\begin{eqnarray}
( \partial_{(2)} E_{3} -   \partial_{(3)} E_{2})+ (v_{20} E_{3}
v_{30} E_{2} ) \nonumber
\\
+ [-(p_{22}+p_{33}) E_{1} + p_{12} E_{2} + p_{13} E_{3} ]
\nonumber
\\
+
 \partial_{(0)} B _{1} +(p_{20}  B_{3} - p_{30}  B_{2}  )
 \nonumber
 \\
 -
[-(v_{22}+v_{33}) B_{1} + v_{12} B_{2} + v_{13} B_{3}]=0 \; ,
\label{11.11a}
\end{eqnarray}
\begin{eqnarray}
( \partial_{(2)} B_{3} -   \partial_{(3)} B_{2})+ (v_{20} B_{3} -
v_{30} B_{2} ) \nonumber
\\
+ [-(p_{22}+p_{33}) B_{1} + p_{12} B_{2} + p_{13} B_{3} ]
\nonumber
\\
-
 \partial_{(0)} E _{1} -(p_{20} E_{3} - p_{30} E_{2}  )
 \nonumber
 \\
 +
[-(v_{22}+v_{33}) E_{1} + v_{12} E_{2} + v_{13} E_{3}]= {1 \over
\epsilon_{0}} j^{1} \; , \label{11.11b}
\end{eqnarray}
\begin{eqnarray}
( \partial_{(3)} E_{1} -   \partial_{(1)} E_{3})+ (v_{30} E_{1} -
v_{10} E_{3} )\nonumber
\\
 + [ p_{21} E_{1} -(p_{11}+p_{33}) E_{2}   + p_{23}
E_{3} ]  \nonumber
\\
+
 \partial_{(0)} B _{2} +(p_{30}  B_{1} - p_{10}  B_{1}  )
 \nonumber
 \\
 -
[ v_{21} B_{1} -(v_{11}+v_{33}) B_{2}  + v_{23} B_{3}]=0 \; ,
\label{11.12a}
\end{eqnarray}
\begin{eqnarray}
( \partial_{(3)} B_{1} -   \partial_{(0)} B_{3})+ (v_{30} B_{1} -
v_{10} B_{3} ) \nonumber
\\ + [ + p_{31} B_{1}   -(p_{11}+p_{33}) B_{2} +
p_{23} cB_{3}]  \nonumber
\\
-
 \partial_{(0)} E _{2} -(p_{30} E_{1} - p_{10} E_{3}  )
 \nonumber
 \\
 +
[ v_{21} E_{1}    -(v_{11}+v_{33}) E_{2} + v_{23} E_{3} ]= {1
\over \epsilon_{0}} j^{2} \; , \label{11.12b}
\end{eqnarray}
\begin{eqnarray}
( \partial_{(1)} E_{2} -   \partial_{(2)} E_{1})+ (v_{10} E_{2} -
v_{20} E_{1} ) \nonumber
\\
+ [ p_{31} E_{1}    + p_{32} E_{2} -(p_{11}+p_{22}) E_{3}  ]
\nonumber
\\
+
 \partial_{(0)} B _{3} +(p_{10}  B_{2} - p_{0}  B_{1}  ) \nonumber
 \\
 -
[ v_{31} B_{1} + v_{32} B_{2} -(v_{11}+v_{22}) B_{3}  ]=0 \; ,
\label{11.13a}
\\
(\partial_{(1)} B_{2} -   \partial_{(2)} B_{1}) + (v_{10} B_{2} -
v_{20} B_{1} ) \nonumber
\\
+ [ + p_{31} B_{1}    + p_{32} B_{2} -(p_{11}+p_{22}) B_{3} ]
\nonumber
\\
-
 \partial_{(0)} E _{3} -(p_{10} E_{2} - p_{20} E_{1}  )
 \nonumber
 \\
 +
[ v_{31} E_{1}  + v_{32} E_{2}  -(v_{11}+v_{22}) E_{3} ]= {1 \over
\epsilon_{0}} j^{3} \; . \label{11.13b}
\end{eqnarray}

\noindent We have obtained  rather  complicated system of eight
equations, in next Section we will prove its equivalence to tensor
generally covariant Maxwell equations.

\section{Relations  between matrix and tensor Maxwell  equations
 }

In the generally covariant tensor Maxwell equations
\begin{eqnarray}
\nabla^{\alpha} F^{\beta \gamma} +  \nabla^{\beta} F^{\gamma
\alpha} +  \nabla^{\gamma} F^{\alpha \beta} = 0 \; , \qquad
\nabla_{\beta }  F^{\beta \alpha } = {1 \over \epsilon_{0}} \;
j^{\alpha} \label{12.1}
\end{eqnarray}

\noindent let us  introduce tetrad field variables, then they take
the form
\begin{eqnarray}
     \partial_{(n)} F_{(m)(l)}   +\gamma_{mbn}      \;   F^{(b)}_{\;\;(l)}
  - \gamma_{lbn}   \;    F_{\;\; (m)}^{(b)}
\nonumber
\\
+ \;    \partial_{(m)}\; F_{(l)(n)} \;  + \gamma_{lbm}
  \;    F^{(b)}_{\;\;(n)}   -\gamma_{nbm}     \;  F_{\;\;(l)}^{(b)}
\nonumber
\\
 +\;  \; \partial_{(l)}F_{(n)(m)} +\gamma_{nbl}    \;    F^{(b)}_{\;\;(m)}  -
  \gamma_{mbl}      \;  F_{\;\;(n)}^{(b)} = 0 \; ,
\label{12.3a}
\\
     \partial_{(b) }   F^{(b)}_{\;\;(c) }
  + e^{\beta}_{(b) ; \beta}  \;   F^{(b)}_{\;\;(c) }    +\gamma_{cab}
   \;   F^{(b)(a)}  =  {1 \over \epsilon_{0} } \;  j_{(c)} \; .
\label{12.3b}
\end{eqnarray}

\noindent Now  we are to detail eqs.  (\ref{12.3a}) and
(\ref{12.3b})   at
\begin{eqnarray}
n, m, l = 1,2,3,  \;\; 0,2,3,  \;\; 0,3,1, \qquad  0, 1, 2 \;;
\;\; \;  \mbox{and} \;\;\; c = 0\;,\; 1\;, \;  2\;, \; 3\; .
\nonumber
\end{eqnarray}

Let it be $ n, m, l = 1,2,3$:
\begin{eqnarray}
     \partial_{(1)} F_{(2)(3)}   +\gamma_{2b1}      \;   F^{(b)}_{\;\;(3)}
  - \gamma_{3b1}   \;    F_{\;\; (2)}^{(b)}
\nonumber
\\
+    \partial_{(2)}\; F_{(3)(1)} \;  + \gamma_{3b2}
  \;    F^{(b)}_{\;\;(1)}   -\gamma_{1b2}     \;  F_{\;\;(3)}^{(b)} \
\nonumber
\\
 +
  \partial_{(3)}F_{(1)(2)} +\gamma_{1b3}    \;    F^{(b)}_{\;\;(2)}  -
  \gamma_{2b3}      \;  F_{\;\;(1)}^{(b)}  = 0 \; ,
\nonumber
\end{eqnarray}

\noindent or
\begin{eqnarray}
    \partial_{(1)} F_{(2)(3)}
+ \gamma_{201}F^{(0)}_{\;\;(3)} + \gamma_{211}F^{(1)}_{\;\;(3)} -
 \gamma_{301} F_{\;\; (2)}^{(0)}  -\gamma_{311} F_{\;\; (2)}^{(1)}     \;
  \nonumber
\\
+     \partial_{(2)}\; F_{(3)(1)} \;  + \gamma_{302}
F^{(0)}_{\;\;(1)} + \gamma_{322} F^{(2)}_{\;\;(1)}
 -\gamma_{102} F_{\;\;(3)}^{(0)} - \gamma_{122} F_{\;\;(3)}^{(2)} \;)
 \nonumber
\\
 +\;  \partial_{(3)}F_{(1)(2)} +
 \gamma_{103} F^{(0)}_{\;\;(2)} + \gamma_{133} F^{(3)}_{\;\;(2)}
  -
  \gamma_{203}   F_{\;\;(1)}^{(0)} -  \gamma_{233}   F_{\;\;(1)}^{(3)}  = 0 \;
\nonumber
\end{eqnarray}

\noindent which with  notation
\begin{eqnarray}
(F_{(2)(3)} , F_{(3)(1)}, F_{(1)(2)}) =( cB_{(i)})  \; , \qquad
(F_{(0)(1)} , F_{(0)(2)}, F_{(0)(3)} =( E_{(i)}) \nonumber
\end{eqnarray}

\noindent reads (again let $c=1$)
\begin{eqnarray}
 - \partial_{(k)}  B_{(k)} +
[\;  (p_{23} -p_{32}) B_{(1)}  +  (p_{31}-p_{13}) B_{(2)}  +
(p_{12}-p_{21}) B_{(3)} \; ]   \nonumber
\\
- [\;  (v_{23} -v_{32}) E_{(1)}  +  (v_{31}-v_{13}) E_{(2)}  +
(v_{12}-v_{21}) E_{(3)} \; ]=  0\; , \nonumber
\end{eqnarray}

\noindent which  coincides with eq. (\ref{11.10b}), if
\begin{eqnarray}
E_{k}= E_{(k)}= {\bf E}  \; , \qquad B_{k} = - B_{(k)} ={\bf B} \;
. \label{12.4c}
\end{eqnarray}

 Let it be  $ n, m, l = 0,1,2 : $
\begin{eqnarray}
   \partial_{(0)} F_{(1)(2)}   +\gamma_{1b0}      \;   F^{(b)}_{\;\;(2)}
  - \gamma_{2b0}   \;    F_{\;\; (1)}^{(b)}
\nonumber
\\
+ \;     \partial_{(1)}\; F_{(2)(0)} \;  + \gamma_{2b1}
  \;    F^{(b)}_{\;\;(0)}   -\gamma_{0b1}     \;  F_{\;\;(2)}^{(b)}
\nonumber
\\
 +
   \partial_{(2)}F_{(0)(1)} +\gamma_{0b2}    \;    F^{(b)}_{\;\;(1)}  -
  \gamma_{1b2}        F_{\;\;(0)}^{(b)} = 0 \; ,
\label{12.5a}
\end{eqnarray}

\noindent and further
\begin{eqnarray}
    \partial_{(0)}c B_{(3)}  -v_{10}    E_{(2)}  -p_{20}  c B_{(1)}
  +v_{20}  E_{(1)}  + p_{10}  cB_{(2)}
\nonumber
\\
  -   \; \partial_{(1)}\; E_{(2)} \;  - p_{31} E_{(1)}  + p_{11}  E_{(3)}
 + v_{11} c B_{(3)}   - v_{31} cB_{(1)}
\nonumber
\\
 +     \; \partial_{(2)}E_{(1)} +
 v_{22}  c B_{(3)} - v_{32}  c  B_{(2)} -
  p_{32}   E_{(2)} + p_{22}  E_{(3)}     = 0 \; ,
\nonumber
\end{eqnarray}

\noindent which coincides with (\ref{11.13a}) multiplied by $ -1$.

  Let  it be $c=0$ in (\ref{12.3b}):
\begin{eqnarray}
     \partial_{(b) }   F^{(b)}_{\;\;(0) }
  + e^{\beta}_{(b) ; \beta}  \;   F^{(b)}_{\;\;(0) }    +\gamma_{0ab}
   \;   F^{(b)(a)}  =  {1 \over \epsilon_{0} } \; \rho \; .
\label{12.6a}
\end{eqnarray}

\noindent Allowing for the identity
\begin{eqnarray}
e^{\beta}_{(b) ; \beta}  \;   F^{(b)}_{\;\;(0) } = -
\gamma_{kc}^{\;\;\;\;c}    F^{(k)}_{\;\;(0) }=
 -(\gamma_{k00} - \gamma_{k11} -
\gamma_{k22} - \gamma_{k33}) F^{(k)}_{\;\;((0) }  \nonumber
\\
= - \gamma_{k00} F^{(k)}_{\;\;(0) } + \gamma_{211}
F^{(2)}_{\;\;(0) }  + \gamma_{311} F^{(3)}_{\;\;(0) } \nonumber
\\
+ \gamma_{122} F^{(1)}_{\;\;(0) }+ \gamma_{322} F^{(3)}_{\;\;(0) }
+\gamma_{133} F^{(1)}_{\;\;(0) }+ \gamma_{233} F^{(2)}_{\;\;(0) }
\; , \nonumber
\end{eqnarray}

\noindent we get
\begin{eqnarray}
    \partial_{(k) }   E_{(k)}
 - p_{31} E_{(2)} + p_{21} E_{(3)}
+ p_{32} E_{(1)} - p_{12} E_{(3)} -p_{23} E_{(1)} + p_{13} E_{(2)}
- \nonumber
\\
  - v_{12} B_{(3)}+ v_{13}B_{(2)}+
 v_{21} B_{(3)} - v_{23} B_{(1)}
 -  v_{31} B_{(2)} + v_{32} B_{(1)}  =  {1 \over \epsilon_{0} } \; \rho \; ,
\nonumber
\end{eqnarray}

\noindent the latter coincides  with (\ref{11.11a}).

Now, let $c=3$ in (\ref{12.3b}):
\begin{eqnarray}
     \partial_{(b) }   F^{(b)}_{\;\;(3) }
  + e^{\beta}_{(b) ; \beta}  \;   F^{(b)}_{\;\;(3) }    +\gamma_{3ab}
   \;   F^{(b)(a)}  =  {1 \over \epsilon_{0} } \;  j_{(3)} \; ,
\label{12.7a}
\end{eqnarray}

\noindent from whence it follows
\begin{eqnarray}
    - \partial_{(0) }   E_{(3) } -  \partial_{(1) }  B_{(2)}   + \partial_{(2) }  B_{(1)}
\nonumber
\\
 - v_{10}  B_{(2)}  +  v_{20} B_{(1)} - v_{11} E_{(3)} -
  p_{31}  B_{(1)} -   v_{22} E_{(3)} -  p_{32} B_{(2)}
\nonumber
\\
+v_{31}    E_{(1)} + v_{32}       E_{(2)}  +
  p_{20}   E_{(1)} + p_{22}    B_{(3)} -
 p_{10}  E_{(2)} + p_{11}      B_{(3)}
 =  -{1 \over \epsilon_{0} } \;  j_{(3)}\; ,
\nonumber
\end{eqnarray}

\noindent the latter  coincides with  (\ref{11.13b}) multiplied by
$(-1)$. In the same manner one  can verify all remaining
equations. Thus, the  matrix and tensor forms of the  Maxwell
equations are equivalent to each other:
\begin{eqnarray}
\alpha^{\alpha}(x) \; [\; \partial_{\rho}  + A_{\alpha}(x)\; ]  \;
\Psi = J(x)  \nonumber
\\
\nabla^{\alpha} F^{\beta \gamma} +  \nabla^{\beta} F^{\gamma
\alpha} +  \nabla^{\gamma} F^{\alpha \beta} = 0 \; , \qquad
\nabla_{\beta }  F^{\beta \alpha } = {1 \over \epsilon_{0}} \;
j^{\alpha} \; . \label{12.8b}
\end{eqnarray}

\section{ Relations  between matrix and tensor   equations   in  media
 }

Let us find detailed tetrad component form for generally covariant
matrix Maxwell equation in
 presence of a  media:
\begin{eqnarray}
\alpha_{\rho}(x) ( \partial_{\rho} + A_{\rho} ) \;M  +
\beta_{\rho}(x) ( \partial_{\rho} + B_{\rho} ) \;N  = J \nonumber
\\
M= \left | \begin{array}{c} 0 \\ {\bf M} \end{array} \right |,
\qquad N= \left | \begin{array}{c} 0 \\ {\bf N} \end{array} \right
|, \qquad  J = {1 \over  \epsilon_{0} \epsilon} \left |
\begin{array}{c} \rho \\ i\;{\bf j} \end{array} \right |
\nonumber
\\
{\bf M} =   { {\bf h } + {\bf f}  \over 2 }  = {1 \over 2}({{\bf
D} \over \epsilon_{0}} + {\bf E}) + {i \over 2}(c{\bf B} + {{\bf
H}\over \epsilon_{0} c})  \nonumber
\\
{\bf N} =  { {\bf h }^{*} - {\bf f}^{*} \over 2 }  = {1 \over 2}({
{\bf D} \over \epsilon_{0}} - {\bf E}) + {i \over 2}(c{\bf B} -
{{\bf H} \over \epsilon_{0}c})\; . \label{13.1}
\end{eqnarray}

\noindent For a time we will use shortening notation:
\begin{eqnarray}
{{\bf D} \over \epsilon_{0}} \Longrightarrow {\bf D} \; , \qquad
c{\bf B}  \Longrightarrow {\bf B}\; , \qquad {{\bf H} \over
\epsilon_{0}c} \Longrightarrow {\bf H} \; . \nonumber
\end{eqnarray}

\noindent Eq. (\ref{13.1}) can be rewritten as follows:
\begin{eqnarray}
-i   (  e_{(0)}^{\rho} \partial_{\rho}  +  {1 \over 2} j^{ab}
\gamma_{ab0}  )M + \alpha^{k}  (  e_{(k)}^{\rho}
\partial_{\rho}  + {1 \over 2} j^{ab} \gamma_{abk} \; )M
\nonumber
\\
-i   (  e_{(0)}^{\rho} \partial_{\rho}  +  {1 \over 2} j^{ab}
\gamma_{ab0}  )N + \beta^{k}  (  e_{(k)}^{\rho}
\partial_{\rho}  + {1 \over 2} j^{*ab} \gamma_{abk} \; )N
= J(x)\; . \label{13.4}
\end{eqnarray}

\noindent Eq. (\ref{13.4}) can be transformed to the form
\begin{eqnarray}
-i   [   \partial_{(0)}  + {\bf s} ({\bf p}_{0} +i{\bf v}_{0}
 ) ] M + \alpha^{k}  [   \partial_{(k)}  + {\bf s}
({\bf p}_{k} +i{\bf v}_{k}  ) ]  M \;  \nonumber
\\
-i   [   \partial_{(0)}  + {\bf s} ({\bf p}_{0} -i{\bf v}_{0}
 ) ] N + \beta^{k}  [ \partial_{(k)}  + {\bf s}
({\bf p}_{k} -i{\bf v}_{k} )  ]  N = J(x)\; . \nonumber
\end{eqnarray}

\noindent Let us divide  it  into real and imaginary parts:
\begin{eqnarray}
(  \alpha^{k}    \partial_{(k)}  + {\bf s} {\bf v}_{0}  +
 \alpha^{k}  {\bf s} {\bf p}_{k}  ){1 \over 2} \left | \begin{array}{c}
0 \\ {\bf D} +{\bf E}
\end{array} \right |
\nonumber
\\
+
 (   \partial_{(0)}   +
  {\bf s} {\bf p}_{0}  - \alpha^{k} {\bf s} {\bf v}_{k}) {1\over 2} \left | \begin{array}{c}
0 \\ {\bf B} + {\bf H}
\end{array} \right |
\nonumber
\\
+ (   \beta^{k}    \partial_{(k)}  - {\bf s} {\bf v}_{0}  +
 \beta^{k}  {\bf s} {\bf p}_{k}  )  {1 \over 2} \left | \begin{array}{c}
0 \\ {\bf D} -{\bf E}
\end{array} \right |
 \nonumber
 \\
  + (   \partial_{(0)}
  {\bf s} {\bf p}_{0}  + \beta^{k} {\bf s} {\bf v}_{k})  {1 \over 2}\left | \begin{array}{c}
0 \\ {\bf B} -{\bf H}
\end{array} \right |
 = {1 \over \epsilon_{0}} \left | \begin{array}{c}
 \rho  \\   0
 \end{array} \right |  .
\label{13.6}
\end{eqnarray}

\begin{eqnarray}
(  \alpha^{k}   \partial_{(k)}  + {\bf s} {\bf v}_{0}
 +
 \alpha^{k}  {\bf s} {\bf p}_{k}  )  {1 \over 2}  \left | \begin{array}{c}
0 \\ {\bf B} \nonumber
\\
+{\bf H}
\end{array} \right |
\nonumber
\\
-
 (  \partial_{(0)}   +
  {\bf s} {\bf p}_{0}  - \alpha^{k} {\bf s} {\bf v}_{k})  {1 \over 2} \left | \begin{array}{c}
0 \\ {\bf D} +{\bf E}
\end{array} \right |
\nonumber
\\
+ (\;  \beta^{k}   \partial_{(k)}  - {\bf s} {\bf v}_{0}  +
 \beta^{k}  {\bf s} {\bf p}_{k}  ) {1 \over 2} \left | \begin{array}{c}
0 \\ {\bf B} -{\bf H}
\end{array} \right |
\nonumber
\\
-
 (  \partial_{(0)}   +
  {\bf s} {\bf p}_{0}  + \beta^{k} {\bf s} {\bf v}_{k}) {1 \over 2}  \left | \begin{array}{c}
0 \\ {\bf D} -{\bf E}
\end{array} \right |
 = {1 \over \epsilon_{0}} \left | \begin{array}{c}
 0  \\   {\bf j}
 \end{array} \right |  .
\label{13.7}
\end{eqnarray}

\noindent From these one can derive the following explicit
equations:

\noindent Let us detail eqs. (\ref{13.6}) -- we will specify only
two cases:
\begin{eqnarray}
 \partial_{(k)} D_{k} -
  (p_{23} -p_{32}) D_{1}   -  (p_{31}-p_{13})D_{2}  -  (p_{12}-p_{21}) D_{3}
\nonumber
\\
+  (v_{23} -v_{32})  H_{1}   +  (v_{31}-v_{13}) H_{2}   +
(v_{12}-v_{21})  H_{3}
 =  \rho \; ,
\nonumber
\\
 \partial_{(2)} E_{3} -  \partial_{(3)} E_{2}
 +
 v_{20}  E_{3} - v_{30}  E_{2}
  -
(p_{22}+p_{33})E_{1} + p_{12} E_{2} + p_{13} E_{3} + \nonumber
\\
p_{20} B_{3}  - p_{30} B_{2} + (v_{22}+v_{33}) B_{1} - v_{12}B_{2}
- v_{13}B_{3} =0 \; . \nonumber
\end{eqnarray}

\noindent Now let us consider two equations from (\ref{13.7}):
\begin{eqnarray}
 \partial_{(k)}  B_{k}   -
  (p_{23} -p_{32}) cB_{1}  -  (p_{31}-p_{13})  B_{2}-
(p_{12}-p_{21})  B_{3} \nonumber
\\
-
  (v_{23} -v_{32})  E_{1}  -  (v_{31}-v_{13}) E_{2}  -
(v_{12}-v_{21})  E_{3}= 0 \nonumber
\\
 \partial_{(2)} H_{3}   -  \partial_{(3)} H_{2}
+
 v_{20}  H_{3}   - v_{30}  H_{2} -(p_{22}+p_{33})  H_{1}  + p_{12}
 H_{2}   + p_{13} H_{3}
\nonumber
\\
- p_{20} D_{3} + p_{30} D_{2} - (v_{22}+v_{33})  D_{1}  +
v_{12}D_{2}  + v_{13}D_{3} = j^{1}\; . \nonumber
\end{eqnarray}

Evidently, these equations  (and their  cyclic counterparts) are
equivalent to tensor generally covariant Maxwell equations
\begin{eqnarray}
\nabla^{\alpha} F^{\beta \gamma} +  \nabla^{\beta} F^{\gamma
\alpha} +  \nabla^{\gamma} F^{\alpha \beta} = 0 \; , \qquad
\nabla_{\beta }  H^{\beta \alpha } =  \; j^{\alpha} \label{13.13}
\end{eqnarray}

\noindent in tetrad representation
\begin{eqnarray}
(F_{(2)(3)} , F_{(3)(1)}, F_{(1)(2)}) =( cB_{(i)})  \; , \qquad
F_{(0)(i)}  =  E_{(i)} \; \nonumber
\\
(H_{(2)(3)} , H_{(3)(1)}, H_{(1)(2)}) =( H_{(i)} /  c )  \; ,
\qquad H_{(0)(i)}  =  cD_{(i)}  \; . \nonumber
\end{eqnarray}

\subsection*{Acknowledgements}

This  work was  supported  by Fund for Basic Research of Belarus
F07-314.

Authors are grateful to  Kurochkin Ya.A.  and Tolkachev E.A. for
discussion and advice.

\end{document}